\documentclass[preprint,12pt]{elsarticle}

\usepackage{amssymb}

\usepackage{tabularx}
\newcolumntype{Y}{>{\centering\arraybackslash}X}
\usepackage{subcaption}
\usepackage{color,xcolor,colortbl}

\newcommand{\rev}[1]{{{\color{blue}#1}}{}}
\renewcommand{\rev}[1]{#1}

\def\tsc#1{\csdef{#1}{\textsc{\lowercase{#1}}\xspace}}
\tsc{WGM}
\tsc{QE}

\usepackage{listings}
\usepackage{arydshln}
\usepackage[hidelinks]{hyperref}
\usepackage{amsmath}
\newcommand{\qt}[1]{\textit{``#1''}}

\usepackage{float}
\newfloat{lstfloat}{htbp}{lop}
\floatname{lstfloat}{Listing}

\usepackage{scalerel}
\usepackage{tikz}
\usetikzlibrary{svg.path}

\definecolor{orcidlogocol}{HTML}{A6CE39}
\tikzset{
  orcidlogo/.pic={
    \fill[orcidlogocol] svg{M256,128c0,70.7-57.3,128-128,128C57.3,256,0,198.7,0,128C0,57.3,57.3,0,128,0C198.7,0,256,57.3,256,128z};
    \fill[white] svg{M86.3,186.2H70.9V79.1h15.4v48.4V186.2z}
                 svg{M108.9,79.1h41.6c39.6,0,57,28.3,57,53.6c0,27.5-21.5,53.6-56.8,53.6h-41.8V79.1z M124.3,172.4h24.5c34.9,0,42.9-26.5,42.9-39.7c0-21.5-13.7-39.7-43.7-39.7h-23.7V172.4z}
                 svg{M88.7,56.8c0,5.5-4.5,10.1-10.1,10.1c-5.6,0-10.1-4.6-10.1-10.1c0-5.6,4.5-10.1,10.1-10.1C84.2,46.7,88.7,51.3,88.7,56.8z};
  }
}

\newcommand\orcidicon[1]{\href{https://orcid.org/#1}{\mbox{\scalerel*{
\begin{tikzpicture}[yscale=-1,transform shape]
\pic{orcidlogo};
\end{tikzpicture}
}{|}}}}

\newcommand{\orcid}[1]{\orcidicon{#1}~\href{https://orcid.org/#1}{#1}}

\journal{Journal of Systems and Software}

\begin{document}

\begin{frontmatter}



\title{Experimental evaluation of architectural software performance design patterns in microservices}

\author[gssi,liu]{Willem Meijer}
\ead{willem.meijer@liu.se}

\author[gssi]{Catia Trubiani}
\ead{catia.trubiani@gssi.it}

\author[monash]{Aldeida Aleti}
\ead{aldeida.aleti@monash.edu}

\tnotetext[orcid]{\textit{Orcid:}
    \orcid{0000-0001-8482-3917} (Willem Meijer)\\
    \orcid{0000-0002-7675-6942} (Catia Trubiani)\\
    \orcid{0000-0002-1716-690X} (Aldeida Aleti)
}

\affiliation[gssi]{organization={Gran Sasso Science Institute},
            city={L'Aquila},
            country={Italy}}

\affiliation[liu]{organization={Linköping University},
            city={Linköping},
            country={Sweden}}

\affiliation[monash]{organization={Monash University},
            city={Melbourne},
            country={Australia}}

\begin{abstract}
    Microservice architectures and design patterns enhance the development of large-scale applications by promoting flexibility.
    Industrial practitioners perceive the importance of applying architectural patterns but they struggle to quantify their impact on system quality requirements. Our research aims to quantify the effect of design patterns on system performance metrics, e.g., service latency and resource utilization, even more so when the patterns operate in real-world environments subject to heterogeneous workloads.
    We built a cloud infrastructure to host a well-established benchmark system that represents our test bed, complemented by the implementation of three design patterns: Gateway Aggregation, Gateway Offloading, Pipe and Filters.
    Real performance measurements are collected and compared with model-based predictions that we derived as part of our previous research, thus further consolidating the actual impact of these patterns.
    Our results demonstrate that, despite the difficulty to parameterize our benchmark system, model-based predictions are in line with real experimentation, since the performance behaviors of patterns, e.g., bottleneck switches, are mostly preserved.
    In summary, this is the first work that experimentally demonstrates the performance behavior of microservices-based architectural patterns.
    Results highlight the complexity of evaluating the performance of design patterns and emphasize the need for complementing theoretical models with empirical data. 
\end{abstract}



\begin{highlights}
\item Building an experimental environment to specify architectural design patterns
\item Enabling a concrete and realistic setup for the performance evaluation of patterns
\item Running experiments for \textit{Gateway Aggregation}, \textit{Gateway offloading}, and \textit{Pipes and Filters}
\item Collecting performance measurements and comparison w.r.t. model-based predictions
\item Analyzing the measurements to assess the performance evaluation of design patterns 
\end{highlights}

\begin{keyword}
Design Patterns \sep Microservice Architectures \sep  Performance Evaluation 



\end{keyword}

\end{frontmatter}



\section{Introduction}
\label{sec:introduction}

The specification of design patterns finds its root in the work by Martin~\cite{martin2000design} where the main concepts and principles are outlined, even if their adoption in microservice-based systems has been investigated at a later stage~\cite{taibi2018architectural, soldani2018pains}.
Among the various research challenges, a large amount of attention is still devoted to understanding the impact of design patterns on multiple quality attributes given the complexity behind such evaluation, e.g.,~\cite{heinrich2017performance, feitosa2019what, wijerathna2022mining}.

Some empirical studies that explore the stakeholders' perception of design patterns can be found in the literature.
For example, Vale et al.~\cite{vale2022designing} recently reported on industry experts positively or negatively perceiving the adoption of \rev{cloud} design patterns when evaluating the quality attributes of microservice-based systems.
\rev{The specification of cloud design patterns is tied with the overarching goal of building reliable, scalable, and secure applications in the cloud.
Our focus in cloud development is on the design and implementation of design patterns since these two phases may largely impact the quality of cloud-hosted applications~\cite{microsoft2022cloud}.} 
There exist some approaches in the literature that focus on specific quality attributes, for instance, the performance evaluation of microservice-based architectures is gaining increasing attention from the research community, e.g.,~\cite{li2021understanding, lira2023architecture, zhou2023revisiting, henning2024benchmarking}.
However, to the best of our knowledge, existing works lack a systematic and experimental evaluation that quantitatively estimates whether the design patterns address the quality (performance) issues that practitioners experience.

One of the findings provided by Vale et al.~\cite{vale2022designing} is that there exists a need for quantitative approaches supporting practitioners in the evaluation of design patterns, 
even more so when managing heterogeneous workloads~\cite{chang2018effective}, and this constitutes the motivation for our research.
In our previous work~\cite{pinciroli_performance_2023} we proposed performance models to analyze the characteristics of seven design patterns considered relevant for the system performance~\cite{vale2022designing}.
\rev{This manuscript moves a step forward in the direction of evaluating the actual impact of cloud design patterns through experimentation.
To this end, here we present a follow-up research effort consisting of the following main contributions: 
}

\begin{itemize}
\item[-] The development of an experimental environment to specify architectural design patterns for microservice architectures, thus enabling a performance evaluation of design patterns under a real setup;

\item[-] The design of experiments to collect real performance measures for three selected design patterns, specifically: \emph{Gateway Aggregation}, \emph{Gateway Offloading}, and \emph{Pipes and Filters};

\item[-] The analysis of the obtained performance measures by making a comparison with model-based predictions, thus further assessing the soundness of the performance evaluation.

\end{itemize}

\rev{Our goal is to better understand the peculiarities of cloud design patterns by studying their actual impact on the system performance.
To this end, we develop a cloud infrastructure to collect real performance measurements. This introduces two major difficulties:} 
\rev{i) the implementation of the cloud design patterns is not straightforward, and} ii) the specification of the workload cannot strictly follow the numerical values used in the performance models defined in~\cite{pinciroli_performance_2023}\rev{, since software performance models are known to abstract from runtime routines~\cite{bolch_queueing_2006}}.

\rev{The choice of the three selected patterns (\textit{Gateway Aggregation}, \textit{Gateway Offloading}, and \textit{Pipes and Filters}) is motivated by their specification that fits within the developed environment. The other patterns are excluded due to their complexity in the implementation, e.g., they might include underlying technologies, such as a database, whose management introduces non-trivial processes (e.g., data synchronization). More details on our reasoning for selecting three cloud design patterns and their complexity are reported in Section~\ref{sec:experimentation}.  
Moreover, the validation of theoretical performance models presented in~\cite{pinciroli_performance_2023} is performed through extensive experimentation that highlights a correspondence between resource utilization and system response time with respect to model-based performance predictions~\cite{pinciroli_performance_2023}.
}

The remainder of the paper is organized as follows. Section~\ref{sec:related-work} discusses the related work and argues on the novelty of our research.
Section~\ref{sec:experimental_setup} describes the built experimental setup and the implementation of the design patterns.
Section~\ref{sec:analysis} presents the metrics used to compare the experimental results of design patterns with respect to the previous theoretical predictions, along with the main lessons learned and threats to validity.
Section~\ref{sec:conclusion} concludes the paper with a vision of future research directions.

\section{Related work}
\label{sec:related-work}

Our work is motivated by a large body of literature since microservices and cloud design patterns (along with their quality attributes) are of key interest to industrial practitioners~\cite{sousa2021survey}. 

Velasco-Elizondo et al.~\cite{velasco2016knowledge} make use of information extraction techniques and knowledge representation to analyze architectural pattern descriptions regarding the \textit{performance} quality attribute. Di Francesco et al.~\cite{difrancesco2019architecting} also support our focus on evaluating the \textit{performance} characteristics of microservice-based systems since it has been found a relevant quality attribute. 
This is further supported by Li et al.~\cite{li2021understanding} who propose a systematic literature review and identify \textit{performance} as one of the most critical quality attributes when designing microservice applications.
Moreover, Wijerathna et al.~\cite{wijerathna2022mining} present an empirical investigation using Stack Overflow and they remark that \textit{performance} is one of the most discussed quality attributes.
\rev{It is worth remarking that microservices were initially introduced for internet-scale systems, such as Netflix, for which scalability is a major concern~\cite{dragoni_microservices_2017}.
Scalability is commonly related to performance~\cite{bass_software_2021}, however, it differs by the flexibility of a system to change in capacity when demand changes~\cite{dragoni_microservices_2017, bass_software_2021}.
The literature already captures the main drivers and barriers to microservice adoption~\cite{soldani2018pains, di_francesco_migrating_2018, baskarada_architecting_2020}.}

Our research draws inspiration from Vale et al.~\cite{vale2022designing} who study the experience of practitioners on the adoption of design patterns (i.e., \textit{Design and Implementation} category of Azure~\cite{microsoft2022cloud}).
Their findings indicate a need for evidence on the positive and negative impacts of design patterns.
Consequently, it is important to develop methods that measure the quality of design patterns, as we aim to do in this paper.
Soldani et al.~\cite{soldani2018pains} also analyze the industrial experiences on microservice-based applications and point out that performance testing is recognized as one of the most challenging activities.
Cortellessa et al.~\cite{CortellessaPET22} focus on monitoring the performance characteristics of microservice-based systems. Their approach aims to establish relationships between these performance characteristics and architectural models, thereby supporting the proactive identification and prevention of performance issues. This method not only aids in diagnosing current performance bottlenecks but also provides a framework for anticipating and mitigating potential problems, contributing to more robust and efficient microservice architectures.

In terms of methodologies, in the following, we discuss the approaches that seem most closely related to ours, at least to the best of our knowledge. 
Khomh et al.~\cite{khomh2018understanding} investigate a cloud-based application to evaluate the performance and energy consumption of some design patterns.
Differently from our work, the authors consider only one Azure \textit{Design and Implementation} pattern (i.e., \textit{pipes and filters}) and its impact was evident when combined with other patterns only.
Akbulut et al.~\cite{akbulut2019performance} study the performance of three design patterns (i.e., \textit{API gateway, chain of responsibility}, and \textit{asynchronous messaging}) 
by deploying a microservice application in a private virtual environment, and they found that optimal design patterns strongly rely on the considered scenarios.
Kousiouris et al.~\cite{kousiouris2021self} focuses on the \textit{batch request aggregation} pattern to reduce the latency of cloud environments, and other approaches are proposed by Ali et al.~\cite{ali2022optimizing} to process requests through the serverless paradigm.
Amiri et al.~\cite{amiri2021modeling} evaluate the trade-offs between system reliability (with a Bernoulli-based model) and performance (employing a statistical model) of three microservice architectures (i.e., \textit{central entity, sidecar}, and \textit{dynamic routing} patterns), and they conclude that \textit{dynamic routing} shows a better performance compared to centralized solutions.
Long et al.~\cite{long2022evaluating} present an empirical study on the impact of the \textit{queue-based load leveling} pattern and assess its positive effect on the performance of a serverless application, however, one pattern is analyzed only.
Ma et al.~\cite{ma2022servicerank} present a framework to detect anomalies using a calibration mechanism recognizing patterns and eliminating their negative effects by root cause analysis, however, it acts on software services only.

When looking for very recent research efforts, we found several approaches that deal
with the quality of microservice-based systems and their architectural decisions.
\rev{For instance, Pallewatta et al.~\cite{pallewatta2024microfog} present a framework for managing microservices-based IoT applications. 
They evaluate the performance of different deployment strategies to guarantee optimal placement of microservices.}
In comparison, our work focuses on design patterns since they offer a wider set of architectural alternatives, and it has been demonstrated by several works that design changes have a strong impact on system performance \cite{amiri2021modeling, laurent2022mutation}. 
Henning et al.~\cite{henning2024benchmarking} benchmark stream processing frameworks used by microservices to study their scalability. Interestingly the frameworks show considerable differences in the rate at which resources have to be added to cope with increasing load.
This supports our research effort on evaluating heterogeneous workloads, thus contributing to selecting different architectural choices that rely on the expected workload.
Giamattei et al.~\cite{giamattei2024automated} present a framework to identify causal relations in observed chains of failures, thus evidencing microservices to be improved for increasing the reliability of architectures. 
Panahandeh et al.~\cite{panahandeh2024serviceanomaly} also propose an anomaly detection approach for microservice-based systems.
However, the authors make use of distributed traces and profiling metrics to improve software reliability.
\rev{We share the intent to identify ``failures'', however, differently from~\cite{giamattei2024automated, panahandeh2024serviceanomaly} our focus is on the performance-based characteristics of software architectures. Our research is aimed at understanding which microservices may generate long delays in handling users' requests and high resource utilization, thus leading to bottlenecks.}
Nikolaidis et al.~\cite{nikolaidis2024eclipse} propose a framework that supports the development of service-based software solutions for the cloud to achieve reuse of services. 
Interestingly, when listing the features of their framework, the authors discuss the need for assistance in the selection of design patterns. Hence our research may complement this feature by augmenting the information on the possible performance implications that may arise from the selection of design patterns.

In summary, given the importance of evaluating the performance of microservice applications, our research is novel to the extent of quantifying the performance implications of adopting design patterns in a real experimental environment.
This way, we aim to support software architects in understanding the system performance fluctuation of microservice-based applications, as well as in considering design alternatives, i.e., one of the main contributions of this manuscript.


\section{Design of experiments}
\label{sec:experimental_setup}

\rev{This section provides a brief background on the theoretical performance models used in~\cite{pinciroli_performance_2023}. To experimentally evaluate the theoretical implementation of the design patterns, we describe the technical details for 
the cloud infrastructure used to run the experiments, and the system that is deployed on it.}
The replication package is available at Zenodo~\cite{meijer_2024_replication}.

\rev{

\subsection{Theoretical performance models}
\label{sec:theoretical-performance-models}

    This work validates some theoretical performance models introduced by Pinciroli et al~\cite{pinciroli_performance_2023} whose aim is to test performance behaviors of design patterns under heterogeneous loads.
    Performance models rely on ``queuing networks'' (QN)~\cite{lazowska1984quantitative}, i.e., a well-assessed formalism to 
    predict the system performance. 
    %
    In short, QN models include three main entities: (i) \emph{service centers} (composed of a \emph{server} and a \emph{queue}) representing system resources, (ii) \emph{jobs} representing the requests sharing the resources, and (iii) \emph{links} that connect service centers and form the network topology~\cite{lazowska1984quantitative}. 
    A QN can be represented as a direct graph whose nodes are service centers and their connections are represented by the graph edges. Jobs go through the graph's edge set based on the behavior of customers' service requests.
    Since QN models are abstract representations of real systems, 
    the precision of the prediction results must be validated.
    This work considers 
    three of the QN models described by Pinciroli et al.~\cite{pinciroli_performance_2023}, 
    transferring the experimentation to real microservice architectures.
    Our experimental environment
    preserves the behavior of cloud design patterns including the inter-microservice relationships, their performance characteristics, and the messages circulating in the system, thus enabling a comparison with the theoretical models.
    
}

\subsection{Cloud infrastructure}
\label{sec:computing_environment}

\begin{figure}[!t]
        \centering
        \includegraphics[width=0.6\textwidth]{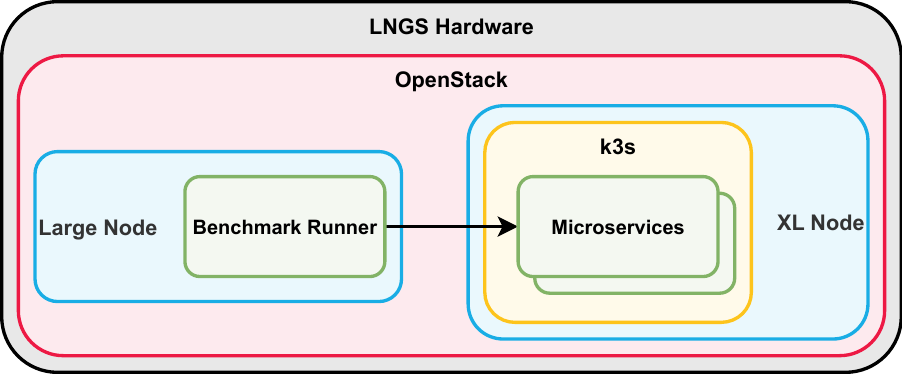}
        \caption{\rev{Overview of the used cluster and experiment topology.}}
        \label{fig:cluster-topology}
    \end{figure}

    \rev{To perform the experiments we set up the small private cloud environment visualized in Figure~\ref{fig:cluster-topology} and explained in the following.}
    To ensure that the experiments are executed in a representative environment, one of the compute clusters at the \textit{Laboratori Nazionali Del Gran Sasso}\footnote{\url{https://www.lngs.infn.it/en}} (LNGS) is used.
    This environment is considered representative because it is intended for infrastructure-as-a-service (IaaS) deployments.
    This system is deployed using multiple hypervisors, each consisting of an \verb|Intel(R) Xeon(R) CPU E5-2697| \verb| v4@ 2.30GHz| CPU, outfitted with 128GB of RAM, and a 1Gb network connection between nodes and towards the outside world.
    This infrastructure is accessed using the OpenStack\footnote{\url{https://www.openstack.org/}} API, an open-source cloud computing solution that provides the functionality to create IaaS deployments.
    This system virtualizes the resources in the physical hardware, making them dynamically accessible by any OpenStack operator.
    Using this framework, a small cluster of two compute nodes is deployed running \verb|Ubuntu-22.04-x86_64-2022-11-13|.
    These compute nodes can be initialized with a configuration representing the amount of CPU cores, RAM, and persistent storage. 
    These configurations can be defined manually.
    However, this study uses standard configurations in OpenStack for simplicity.
    One of these nodes has configuration \verb|large| (i.e., 4 CPU cores, 8GB of RAM, and 40GB of persistent storage), and the second has configuration \verb|xl| (8 CPU cores, 8GB of RAM, and 80GB of persistent storage). 

    To improve the system's flexibility, Kubernetes\footnote{\url{https://kubernetes.io/}} is deployed on top of the OpenStack environment.
    There are many ways in which Kubernetes can be deployed.
    This study used K3S\footnote{\url{https://k3s.io/}}, an open-source Kubernetes distribution that is lightweight by design.
    It should be noted that although the experiments in this study are run using a multi-node cluster, each experiment can be run using Minikube\footnote{\url{https://minikube.sigs.k8s.io/}} as well (i.e., a single-node cluster).
    \rev{The microservices are deployed within the K3S cluster using $\mu$Bench~\cite{detti_bench_2023} (see Section~\ref{sec:microservice-topology-modeling}), and a \qt{benchmark runner} is deployed as a standalone process to generate benchmark requests (see Section~\ref{sec:experimentation}).}

\subsection{Experimental settings}
\label{sec:experimentation}

    This study evaluates three out of seven performance-related design patterns described by Pinciroli et al.~\cite{pinciroli_performance_2023} by performing an independent \rev{validation study~\cite{kitchenham1995towards},} 
    specifically \textit{gateway aggregation}, \textit{gateway offloading}, and \textit{pipes and filters}.
    The reason for this selection is that implementing and testing design patterns is a time-consuming process.
    Pinciroli et al.~\cite{pinciroli_performance_2023} considered four additional patterns: \textit{command query responsibility segregation (CQRS)}, \textit{static content hosting}, \textit{anti-corruption layer}, and \textit{backends for frontends}.
    The first two patterns were not selected because their behaviors strongly depend on implementation considerations and underlying technologies.
    For example, CQRS --- a pattern that separates read requests from write requests in a database by separating their data models --- implements a synchronization step to ensure the different data models have an equivalent state.
    Such a synchronization step highly depends on the system requirements.
    For instance, it has a larger impact on systems that prioritize consistency than on systems that focus on availability and use eventual consistency.
    Consequently, experimental evaluation of these patterns should consider these implementation details thoroughly.
    \rev{According to Pinciroli et al.~\cite{pinciroli_performance_2023}, the \textit{anti-corruption layer} and \textit{backends for frontends} patterns are anticipated to perform similarly under diverse loads to the \textit{gateway offloading} and \textit{gateway aggregation} patterns, offering minimal additional insights for comparison.
    The \textit{anti-corruption layer} --- a pattern that introduces an adapter to translate requests between different systems --- would require the usage of different technologies for the development of the involved systems, complicating the implementation of the adapter component.
     The \textit{backends for frontends} --- a pattern that avoids customizing a single backend microservice for multiple client interfaces --- would imply implementing frontends with different requirements and constraints, introducing complexity in the load generation process.
    Hence, the \textit{anti-corruption layer} and \textit{backends for frontends} were not included in the study.}
    All these experiments are implemented as described in Section~\ref{sec:microservice-topology-modeling}, i.e., using \verb|json| configuration files that can be consumed by $\mu$Bench~\citep{detti_bench_2023}.
    To ease the process of running experiments with different parameters (i.e., heterogeneous workloads) custom code was written that alters the configuration files of $\mu$Bench.

    The experiments are analogous to the ones described in~\cite{pinciroli_performance_2023} (see more details in Section~\ref{sec:microservice-topology-modeling}).
    The systems were tested for heterogeneous workloads, similar to the theoretical models~\cite{pinciroli_performance_2023}, i.e., by shifting request emphasis from one service to another.
    It is worth noticing that this study makes a slight virtual distinction compared to \cite{pinciroli_performance_2023} as they define multiple users each of which sends a specific type of request. 
    In contrast, this study defines a single user having a probability of sending a request type (i.e., 
    we do aggregate the behavior of the separate users in~\cite{pinciroli_performance_2023}).
    Both studies define only two request types $R = \{r_1, r_2\}$ per experiment, for which a heterogeneous load $H$ is any set of requests containing only these messages $H \in R^*$ (i.e., an element of the Kleene closure).
    This study defines multiple heterogeneous loads using the probability of observing either message in the set; such that $p(r_1) = 1 - p(r_2)$. 
    
    It is worth noticing that the level of granularity in this study is intentionally lower compared to our previous work~\cite{pinciroli_performance_2023} where 25 and 50 heterogeneous load distributions were used depending on the experiment.
    This study limits itself to 5 heterogeneous load distributions to cope with the amount of computation time.
    Running the complete experiment with the lower granularity takes over 24 hours.
    Increasing the level of granularity would merely increase computation time and barely increase the quality of the results.
    The used heterogeneous configurations are defined as follows: $p(r_1) \in \{\frac{0}{5}, \frac{1}{5}, ..., \frac{5}{5}\}$.
    These were picked because they are uniformly distributed across all possible heterogeneous loads, thus observing global performance changes in the system.
    Each tested experimental configuration was stressed for 5 minutes at a time, during which the system was tasked to process as many requests as possible.
    Because some system configurations simply perform worse than others, this means that the number of processed messages differs across experiments.
    However, 5 minutes is deemed sufficiently large as, using this limit, every experiment processed at least 9k requests, with an average of almost 15k requests.
    
    After an experiment was completed, information regarding CPU utilization and overall request delay was acquired.
    \rev{
    We are aware that further metrics, such as system scalability and energy consumption, are important when designing microservice architectures~\cite{vale_designing_2022}.
    In this study, we exclude these metrics, as 
    we are interested in validating the theoretical models used by Pinciroli et al.~\cite{pinciroli_performance_2023}, and we collect the very same metrics. Our metrics align with the literature showing an increased interest in service performance as one of the most critical quality attributes in addition to scalability, e.g.,~\cite{wijerathna2022mining, li2021understanding, difrancesco2019architecting}. 
    }
    
    Request delay is a metric natively collected by $\mu$Bench.
    CPU utilization is not, for which this was collected using Prometheus\footnote{\url{https://prometheus.io}}.
    This is a popular system that collects system metrics and was also used by the creators of $\mu$Bench. 
    To collect the relevant data, the raw CPU utilization was queried through the Prometheus API.
    Each collected data point represents the cumulative CPU usage of a service in seconds since it started.
    Because this is cumulative, the discrete derivative ($\frac{u(t + \Delta t) - u(t)}{\Delta t}$) between two consecutive data readings is calculated to estimate the average CPU utilization between those two moments in time.
    For example, given two readings, $u = 15$ at $t = 1s$ and $u = 26$ at $t = 14s$, we know that the average CPU utilization during that period is $\frac{26 - 15}{14 - 1} \approx 0.85~CPU/s$; i.e., $85\%$ CPU utilization.
    \rev{
    It is worth remarking that performance results can be affected by background processes. 
    To mitigate the risk of unreliable results, we follow recommendations made in experimental guidelines~\cite{bartz-beielstein_benchmarking_2020, beiranvand_best_2017, gregg_systems_2014}, 
    executing each experiment multiple times, and using the average results for further analysis.
    We repeated each experiment 6 times that we found, at the experimental stage, an upper bound to avoid infeasibly long experiments.  
    }
    
    We recall that a second compute node was deployed in the OpenStack cluster (see Section~\ref{sec:computing_environment}).
    This node was used during the experimentation phase, such that the load generator process was run on this node.
    The load generator was deployed in this fashion (instead of using separate hardware) because it simplifies the experimental setup and reduces the risk of network issues affecting the experimental results.
    For example, the network connecting two separate systems could have timed out or the response delay could have been affected by increased network activity (e.g., when many other devices or people are using the network).
    This means the load generator and the microservice system ran in physical proximity, which almost certainly decreased the communication delay measured by the experiments.
    A consequence of this is that the experiments somewhat misrepresent reality.
    However, this does not affect the measured results as using separate hardware would only increase the response delay by some constant and some noise.
    Because this study evaluates the performance behaviors of the design patterns (i.e., the relative results) and not absolute differences, removing a constant does not affect drawn conclusions.
    Additionally, removing noise only increases the consistency of the experiments, making it possible to draw clearer results.
    Ultimately then, even though it might slightly misrepresent reality, running the load generator and the microservice deployment in physical proximity improves the quality of results in the best scenario and does not alter them at all in the worst scenario, for which this design decision is deemed acceptable.

\subsection{Microservice modeling}
\label{sec:microservice-topology-modeling}

    The primary goal of this study is to evaluate performance design patterns in microservice architectures.
    In recent years, various benchmarking systems have been developed for this purpose, like TeaStore~\citep{von_kistowski_teastore_2018, eismann_teastore_2019, kounev_teastore_2020} or DeathstarBench~\citep{gan_open-source_2019}, which are manually designed microservice deployments.
    Although these benchmarking systems are relevant in their regard, they offer limited flexibility because they implement concrete systems (i.e., systems with explicit specifications).
    \rev{
    Although these systems offer some configurability, e.g., memory vs. CPU-intensive tasks, fine-tuning such a demand to the level necessary in this study is a difficult and cumbersome process.
    In addition, changing the topology of these systems to implement design patterns is challenging as the 
    correctness of the implementation cannot be guaranteed.}
    
    To tackle this problem, two recent studies~\cite{detti_bench_2023, sedghpour_hydragen_2023} have developed flexible benchmarking systems.
    \rev{Functionally, both these systems are very similar. 
    They allow users to design a deployment topology and service loads through configuration files.
    These configurations are then translated to microservices that can be deployed on a Kubernetes cluster.
    This work uses the $\mu$Bench system proposed by Detti et al.~\cite{detti_bench_2023}.
    Their tool was chosen over HydraGen~\cite{sedghpour_hydragen_2023} because the documentation was perceived more complete when starting our experimentation.}
    Broadly put, $\mu$Bench implements two main functionalities: 1) custom service topologies, and 2) custom service loads.
    It achieves these by translating specifications defined in \texttt{json} files into Kubernetes resources (services, deployments, etc.).
    Users can define the various services they want, what other services they communicate with, how frequently they do that, and how much effort it costs to process a request (which can be specified into compute, I/O, memory, or sleep-intensive tasks). 
    \rev{This flexibility makes it possible to define any topology, using as many design patterns as desired.}
    
    Using this specification, the patterns addressed by Pinciroli et al.~\cite{pinciroli_performance_2023} could be directly translated into a usable Kubernetes deployment.
    For example, the gateway offloading pattern was implemented by defining four services: \verb|gw| (the gateway), \verb|s1|, \verb|s2|, and \verb|s3|, where the gateway calls either \verb|s1| or \verb|s2|, and \verb|s2| calls \verb|s3|.
    A CPU-intensive workload is defined per service: calculating decimal points of $\pi$ proportional to the absolute number of milliseconds spent by the service; i.e., if in our previous work~\cite{pinciroli_performance_2023} a service is specified to spend 12 milliseconds to handle a request, its experimental counterpart will calculate $q$ digits of $\pi$ 12 times where $q$ is some constant used to amplify the workload to a realistic amount.
    Inside $\mu$Bench the number of repeats is referred to as \verb|trials| and the calculated number of digits of $\pi$ is called \verb|range_complexity|. 
    \rev{Listing~\ref{lst:gateway-offloading-specification} provides an excerpt of the specification used to evaluate the gateway offloading pattern.
    Each service has a separate entry, comprising its internal and external behavior; respectively, the computation done and the services requested afterward. Both can be “request dependent” so their behavior changes when a request of a different type is received. For instance, in Listing~\ref{lst:gateway-offloading-specification}, the gateway is the only component showing a “request dependent” external behavior since it relies on s1 or s3 requests. 
    The intuition for this specification is that once any of the services receives a request, they perform some internal computation followed by sending a request to 
    other service(s).}

\definecolor{background}{HTML}{EEEEEE}
\definecolor{json_string}{RGB}{20,105,176}
\colorlet{json_default}{magenta!60!black}
\definecolor{json_number}{RGB}{0,0,0} 
\definecolor{line_color}{RGB}{220, 0, 40}

\lstdefinelanguage{json}{
    basicstyle=\ttfamily\color{json_default},
    sensitive=false,
    numbers=left,
    numberstyle=\scriptsize\color{line_color},
    stepnumber=1,
    numbersep=8pt,
    showstringspaces=false,
    breaklines=true,
    backgroundcolor=\color{background},
    commentstyle=\color{gray},
    string=[s]{"}{"},
    stringstyle=\color{json_string},
    morestring=[b]',
    morestring=[b]",
    moredelim=[s][\color{json_number}]{:}{,},
    literate=
     *{:}{{\textcolor{json_number}{:}}}{1}
      {,}{{\textcolor{json_number}{,}}}{1}
      {\{}{{\textcolor{json_number}{\{}}}{1}
      {\}}{{\textcolor{json_number}{\}}}}{1}
      {[}{{\textcolor{json_number}{[}}}{1}
      {]}{{\textcolor{json_number}{]}}}{1},
    frame=lines,
    rulecolor=\color{black}
}

\newcommand{\mylstcaption}[0]{\rev{An excerpt of the specification for the gateway offloading pattern. 
}}

\begin{lstfloat}[!b]
\begin{lstlisting}[language=json,basicstyle=\footnotesize, label={lst:gateway-offloading-specification}, captionpos=b, caption=\mylstcaption]
{
  "gw": {"internal": {"range_complexity": 250, "trials": 20},
         "external": {
           "s1_request": {"services": ["s1"]},
           "s3_request": {"services": ["s3"]}}},
  "s1": {"internal": {"range_complexity": 250, "trials": 20}},
  "s2": {"internal": {"range_complexity": 250, "trials": 12},
         "external": {"services": ["s3"]}},
  "s3": {"internal": {"range_complexity": 250, "trials": 15}}
}
\end{lstlisting}
\end{lstfloat}

    Off-the-shelf, $\mu$Bench did not support all features needed to perform the experiments.
    The most important of these were: i)~multiple request types (i.e., the load generator could generate multiple requests); ii)~request-dependent service loads (for example, the gateway aggregation experiment defines s1 and s3 requests); iii)~communication with services that were not generated by $\mu$Bench (as a custom implementation of the gateway aggregation was used); iv)~a load generator that runs for a set amount of time and pushes the limits of the system (rather than a generator that sends a set number of requests).
    The first two functionalities could be implemented by adding support for custom headers and implementing a means to forward headers in the services.
    We implemented the other functionalities by updating the generation process for Kubernetes resources and adding a new load generator.

\rev{
\subsection{Workload specification}
\label{sec:load-amplification}
}
    
    \rev{To model the workload of all the services, a load specification is used that is analogous to the parameters defined in~\cite{pinciroli_performance_2023}.}
    Although the theoretical and experimental delays should optimally be precisely the same, a limitation of the $\mu$Bench system is that small workloads (i.e., processing times of mere milliseconds, like Pinciroli et al.~\cite{pinciroli_performance_2023} did) is very hard to specify due to the large overhead the system itself has.
    Some tests were done to synchronize the theoretical and experimental parameters, however, the theoretical load was systematically lower than the experimental system deployed with minimum load.
    Therefore, instead of using a specification with very small workloads, these workloads have been amplified to preserve their respective ratios, thus accurately 
    \rev{validating} the behaviors observed in our previous work~\cite{pinciroli_performance_2023} and making a comparison.
    An immediate consequence of this is that the system's throughput is noticeably lower and the response times are noticeably higher.
    This will, however, not meaningfully affect the outcome of this study, as bottleneck transitions remain the same following the basic principle that the amount of work is equal to the number of requests multiplied by the time necessary to process them.
    If the amount of time necessary to process requests is higher, but the number of requests is lower, the ratio remains the same and thus the CPU utilization remains the same.
    This is analogous to merging multiple requests in the client.
    Although this causes the absolute values of message delay to be substantially higher, it does not affect its behavior.

\subsection{Specification of design patterns}
\label{sec:specification}
    
    This study uses the technology proposed by Detti et al.~\cite{detti_bench_2023}, $\mu$Bench, who created a synthetic and flexible benchmarking system for microservices.
    To use this system, it is necessary to define some configuration files describing the topology of the microservice system and the individual microservices' workloads.
    \rev{The systems deployed in this study follow the specification proposed by Pinciroli et al.~\cite{pinciroli_performance_2023} (see Section~\ref{sec:microservice-topology-modeling}). } 
    In the following, we describe these design patterns' respective specifications. 

    The services share several common specifications.
    The most important of these is the CPU limitations of the services, which is set to 1 CPU by default.
    This is specified as \verb|1000m| in Kubernetes and $\mu$Bench, meaning 1k milli-CPU seconds.
    This means that although the compute node on which the service runs has a higher capacity, the service is virtually limited to this amount.
    This might seem like an arbitrary step that limits the potential performance of a service.
    However, in practice, it increases the control of the system and prevents individual services from thrashing.
    For example, given 3 microservices running on a compute node with 4 CPUs, if each service has a CPU limit of 1 CPU, the CPU can never be overcommitted because $3 \times 1 < 4$.
    This, of course, would not hold if $4+$ services are deployed on that node, however, this is never the case in the run experiments.
    In this study, the maximum number of simultaneously run services is 6, whereas the number of CPU cores is 8, thus mitigating the risk of overcommitment.

    Because this study is specifically interested in the behavior of software patterns under heterogeneous loads, experiments are conducted with different types of requests.
    For example, the gateway aggregation uses \verb|s1_intensive| and \verb|s3_intensive| requests.
    The following sections give a specification of the request types and the amount of work performed by services when a request of that type is received.
    As specified in Section~\ref{sec:microservice-topology-modeling}, each service's workload has been amplified.
    The same workload amplification is used across experiments, $250$, which was chosen through experimentation.
    A relatively high number was picked to ensure bottlenecks exist in the system.

    \subsubsection{Gateway aggregation}
    \label{sec:gateway-aggregator}

        \begin{figure}[!b]
            \centering
            \includegraphics[scale=0.7]{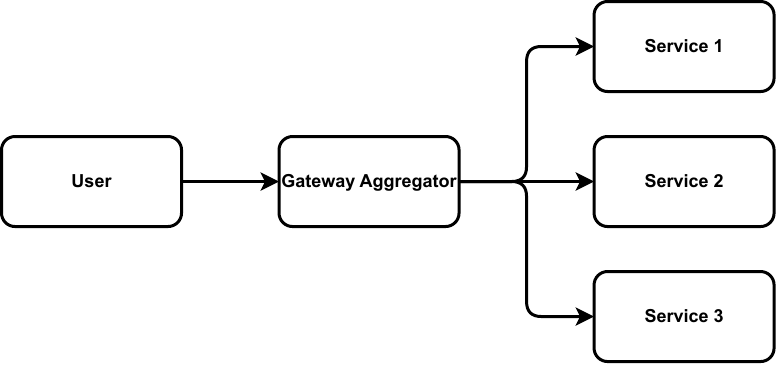}
            \caption{Model of the \textit{gateway aggregation} pattern}
            \label{fig:model-gateway-aggregator}
        \end{figure}
        
        The gateway aggregation pattern\footnote{\url{https://learn.microsoft.com/en-us/azure/architecture/patterns/gateway-aggregation}} is an architectural pattern most commonly applied in scenarios where it is necessary to reduce the number of messages that travel across a network link~\citep{vale_designing_2022}.
        \rev{When inspecting the definition provided by Microsoft~\citep{microsoft2022cloud}, the specification of this pattern indicates using a gateway for aggregating multiple individual requests into a single request. The usefulness of this pattern is evident when a client must make multiple calls to different backend systems to perform an operation.} 
        The main principle of this pattern is that clients combine (aggregate) multiple requests into one and send that as a single request to a gateway aggregator service.
        In turn, this service acts as an ambassador of the client, such that it unwraps the aggregated requests, executes them, aggregates the results into one response, and sends this back to the client.
        \rev{Our interpretation of this pattern abstracts from the network connecting a client and server, given that our focus is on the behavior of the design pattern. From a performance-based perspective, we model and analyze the gateway aggregator component that plays the crucial role of handling aggregated requests and interacting with multiple microservices.
        }

        Figure~\ref{fig:model-gateway-aggregator} visualizes the topology of this pattern, which is equivalent to that used in our previous work~\cite{pinciroli_performance_2023}.
        Here, the three services are implemented using $\mu$Bench.
        However, because $\mu$Bench does not natively support gateway aggregation behavior (i.e., unwrapping aggregated requests and merging their results), a simple custom implementation of this service is used instead\footnote{\url{https://github.com/wmeijer221/muBench-experiment/tree/main/gssi_experiment/gateway_aggregator/gateway_aggregator_service}}.
        A consequence is that no one-to-one comparison can be made between the absolute theoretical and experimental results.
        However, as the gateway is not a system bottleneck (as in~\cite{pinciroli_performance_2023} and confirmed in the results of this work), our custom implementation does not either alter the behavior of the system or the drawn conclusions.

        Table~\ref{tab:workload-gateway-aggregator} gives an overview of the synthetic workloads used by the services.
        This experiment differentiates between two types of messages: \verb|s1_intensive| and \verb|s3_intensive|, differing in the amount of work they demand from either service.
        In this scenario, the runner makes aggregated requests to all microservices, i.e., every user request is processed by all services.
        
        \begin{table}[!t]
        \scriptsize
            \centering
            \begin{tabular}{|c|c|c|}
                \hline
                \textbf{Service} & \textbf{S1 Intensive Requests} & \textbf{S3 Intensive Requests} \\
                \hline\hline
                \textit{Gateway Aggregator} & n/a & n/a \\
                \textit{Service 1} & $18$ & $7$ \\
                \textit{Service 2} & $12$ & $15$ \\
                \textit{Service 3} & $5$ & $20$ \\
                \hline
            \end{tabular}
            \caption{Microservice workload for the \textit{gateway aggregation} pattern.}
            \label{tab:workload-gateway-aggregator}
        \end{table}

    \subsubsection{Gateway offloading}
    \label{sec:gateway-offloading}

        \begin{figure}[!b]
            \centering
            \includegraphics[scale=0.7]{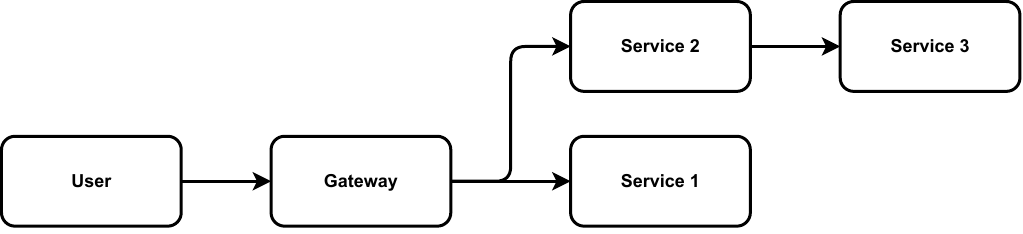}
            \caption{Model of the \textit{gateway offloading} pattern}
            \label{fig:model-gateway-offloading}
        \end{figure}
        
        The gateway offloading pattern\footnote{\url{https://learn.microsoft.com/en-us/azure/architecture/patterns/gateway-offloading}} is a relatively simple pattern that is, among others, used to minimize redundant computation~\citep{ vale_designing_2022}.
        This pattern can be applied by identifying what computation is repeated in multiple services (e.g., authentication, or message decryption), and by extracting (off\-loading) this functionality into a common gateway~\cite{microsoft2022cloud}.
        This way, the gateway executes the common functionality, for which the following services no longer need to concern themselves about it.
        Applied to a fan-out architecture, like the one used in the gateway aggregator experiment, limited benefits might be acquired as computation is merely shifted from one service to another.
        However, applying it to a deeper architecture, like the one tested in this study, might show compounding benefits.
        This is visualized in Figure~\ref{fig:model-gateway-offloading}, where the gateway proxies two services that represent ``pipelines'': the pipeline containing service 1, and the pipeline containing the combination of services 2 and 3.
        In the former, no performance can be acquired from applying this pattern, as the computation performed by service 1 will simply shift to the gateway.
        However, in the latter, computation that would normally be computed by both services 2 and 3 (i.e., two computations) will be shifted into a single computation in the gateway, functionally cutting the necessary computation in half. 

        Table~\ref{tab:workload-gateway-offloading} shows the synthetic workloads used by the various services.
        To limit the computation time necessary to execute the experiments, three off\-load amounts have been selected in this study: $W = \{0, 5, 10\}$ (contrary to the full range $[0, 10]$ adopted by Pinciroli et al.~\cite{pinciroli_performance_2023}).
        Consequently, as the amount of work performed by the gateway changes across experiments, the workloads of the other services are decreased by that amount. 
        For example, if the gateway workload is 5, the workload of Service 1  is $20 - 5 = 15$.
        In this scenario, a request is not processed by every service.
        Instead, the gateway forwards requests to either Service 1 or  Service 2 based on the request type.
        
        \begin{table}[!t]
        \scriptsize
            \centering
            \begin{tabular}{|c|c|c|}
                \hline
                \textbf{Service} & \textbf{Dashboard Requests} & \textbf{Monitoring Requests} \\
                \hline\hline
                \textit{Gateway} & $\{0, 5, 10\}$ & $\{0,5,10\}$ \\
                \textit{Service 1} & $20 - \text{offload}$ & n/a \\
                \textit{Service 2} & n/a & $12 - \text{offload}$ \\
                \textit{Service 3} & n/a & $15 - \text{offload}$ \\
                \hline
            \end{tabular}
            \caption{Microservice workload for the \textit{gateway offloading} pattern.}
            \label{tab:workload-gateway-offloading}
        \end{table}

        \begin{figure}[!b]
            \centering
            \begin{subfigure}{\textwidth}
                \centering
                \includegraphics[scale=0.7]{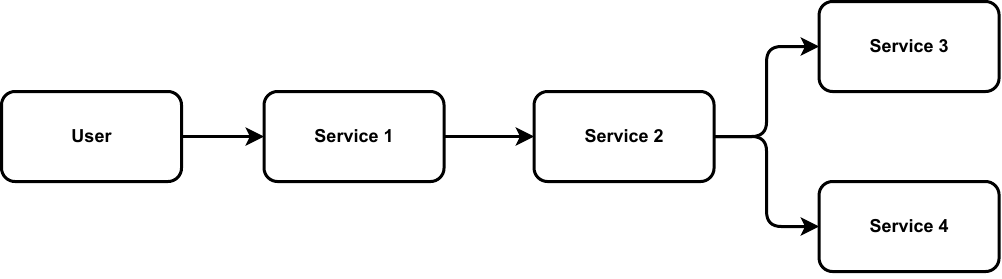}
                \caption{Joint pipeline}
                \label{fig:pipes-and-filters-joint}
            \end{subfigure}
            
            \begin{subfigure}{\textwidth}
                \centering
                \includegraphics[scale=0.7]{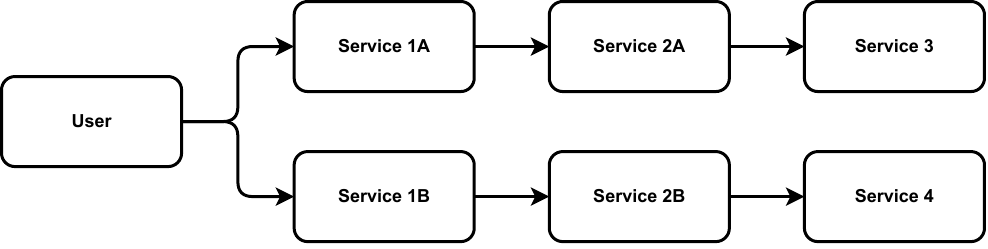}
                \caption{Separated pipeline}
                \label{fig:pipes-and-filters-separated}
            \end{subfigure}
            
            \caption{Models of the \textit{pipes and filters} pattern variants}
            \label{fig:model-pipes-and-filters}
        \end{figure}
        
    \subsubsection{Pipes and filters}
    \label{sec:pipes-and-filters}
    
        The pipes and filters design pattern\footnote{\url{https://learn.microsoft.com/en-us/azure/architecture/patterns/pipes-and-filters}} is a high-level pattern designed to optimize modularity and reusability by splitting functionalities into separate components that can be connected flexibly~\citep{microsoft2022cloud, vale_designing_2022}.
        Similarly to Pinciroli et al.~\cite{pinciroli_performance_2023}, two variants of this pattern have been studied in this work: a pipeline with shared components (shown in Figure~\ref{fig:pipes-and-filters-joint}), and a pipeline with separate resources (shown in Figure~\ref{fig:pipes-and-filters-separated}).
        Since the separated experiment has access to twice the amount of resources, a variant of the joint experiment is used where each shared service has access to 2 CPUs (i.e., a CPU limit of \verb|2000m| in Kubernetes). 
        
        Table~\ref{tab:workload-pipes-and-filters} shows the used service workloads.
        Each experiment uses the same workload; i.e., both Service 2A and Service 2B in the separated model have a workload of 15 for \verb|s3_requests| and 9 for \verb|s4_requests|.
        Because not every service processes each request, some workloads are marked non-applicable.
        
        \begin{table}[!b]
        \scriptsize
            \centering
            \begin{tabular}{|c|c|c|}
                \hline
                \textbf{Service} & \textbf{Service 3 Requests} & \textbf{Service 4 Requests} \\
                \hline\hline
                \textit{Service 1} & $12$ & $8$ \\
                \textit{Service 2} & $15$ & $9$ \\
                \textit{Service 3} & $11$ & n/a \\
                \textit{Service 4} & n/a & $10$ \\
                \hline
            \end{tabular}
            \caption{Microservice workload for the \emph{pipes and filters} pattern.}
            \label{tab:workload-pipes-and-filters}
        \end{table}

\section{Evaluation of analysis results}
\label{sec:analysis}

\subsection{Metrics of interest}
\label{sec:analysis-metrics}

The primary goal of this study is to \rev{validate} 
whether the response time and CPU utilization predicted by the theoretical models in~\cite{pinciroli_performance_2023} hold in real deployments.
To do this, three metrics have been selected that emphasize the relationship between the theoretical and experimental results. 
Because this study is focused on understanding the behavioral trend of design patterns when employed with heterogeneous loads, rather than matching absolute values, Spearman's rank correlation coefficient \cite{sedgwick2014spearman} is used.
\rev{This metric makes it possible to identify the correlation between theoretical and experimental results.
Spearman's rank was chosen instead of Pearson's because the data is not normally distributed.
A statistically significant rank coefficient (i.e., where $p < 0.05$) indicates the strength of the correlation; i.e., the strength of the global relationship between theoretical and experimental results.
These results can be interpreted to see if the change in experimental results across heterogeneous loads approximately follows the same curvature as the theoretical model's.}

To emphasize the deviation between the two datasets, mean average error (MAE) is used (shown in Equation~\ref{eq:mae}), highlighting the average deviation between an observed (experimental) and the expected outcome (synthetical) in absolute numbers.
\rev{Adding MAE to the analysis enables us to complement the global relationship indicated by Spearman's rank with a local comparison between theoretical and experimental results.
Because error can be calculated for individual experiments, this metric highlights whether results differ in specific experiments, even though the behaviors globally match.}

{
    \begin{tabularx}{\linewidth}{YYcc}
        \begin{equation}
            \label{eq:mae}
            MAE = \frac{1}{n}\sum^n_{i = 1} |e_i - o_i|
        \end{equation}
        & 
        \begin{equation}
            \label{eq:min-max-normalization}
            n_Y(x) = \frac{x - \min(\overline{Y})}{\max(\overline{Y}) - \min(\overline{Y})}
        \end{equation}
        &~&~
    \end{tabularx}
}

As described in Section~\ref{sec:load-amplification}, this study makes a trade-off between throughput and request processing time, which increases message response times.
Therefore, the MAE is calculated a second time using a normalized version of the dataset calculated using min-max feature scaling (shown in Equation~\ref{eq:min-max-normalization}).
Although this method is trivially applied to the synthetic dataset, its sensitivity to outliers must be considered when applied to the experimental data.
This is not a problem for CPU utilization data, as this is a percentage, meaning the values are limited within the range $[0, 100]$.
\rev{However, it is a problem for delay data as the range of this variable can reach unexpectedly high values due to environmental factors.
To mitigate this issue, we refrain from using the literal minimum and maximum values measured during the experiments.
Instead, we calculate the average performance across iterations of a single experiment, reducing the impact of outlier values, and calculate the minimum and maximum of the averages across all experiments.
For example, we performed 36 runs that collected data for 6 unique experiments to test the gateway aggregation pattern, each executed 6 times.
We then calculated the average response delay for each experiment, yielding 6 averages.
These were used to calculate the minimum and maximum when normalizing the data.
CPU utilization values were similarly normalized.
}

\subsection{Design patterns results}
\label{sec:results}

In the following, we report the analysis results for the patterns under evaluation, specifically the \textit{gateway aggregation}, the \textit{gateway offloading}, and the \textit{pipes and filters} patterns.

\subsubsection{Gateway aggregation}
\label{sec:results-gateway-aggregator}
    

\begin{figure}
    \centering
    \begin{tabular}{cc}
        \begin{subfigure}{0.5\textwidth}
            \includegraphics[width=\textwidth]{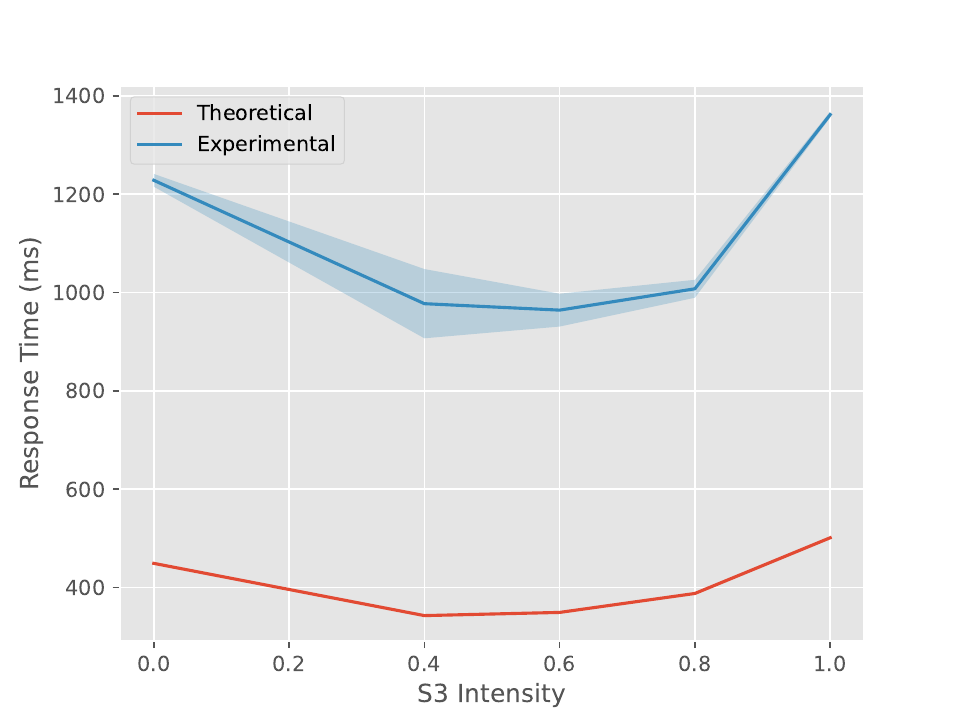}
            \caption{Absolute theoretical~\cite{pinciroli_performance_2023} and experimental request latency}
            \label{fig:ga-req-lat-abs}
        \end{subfigure}
        
        &
        
        \begin{subfigure}{0.5\textwidth}
            \includegraphics[width=\textwidth]{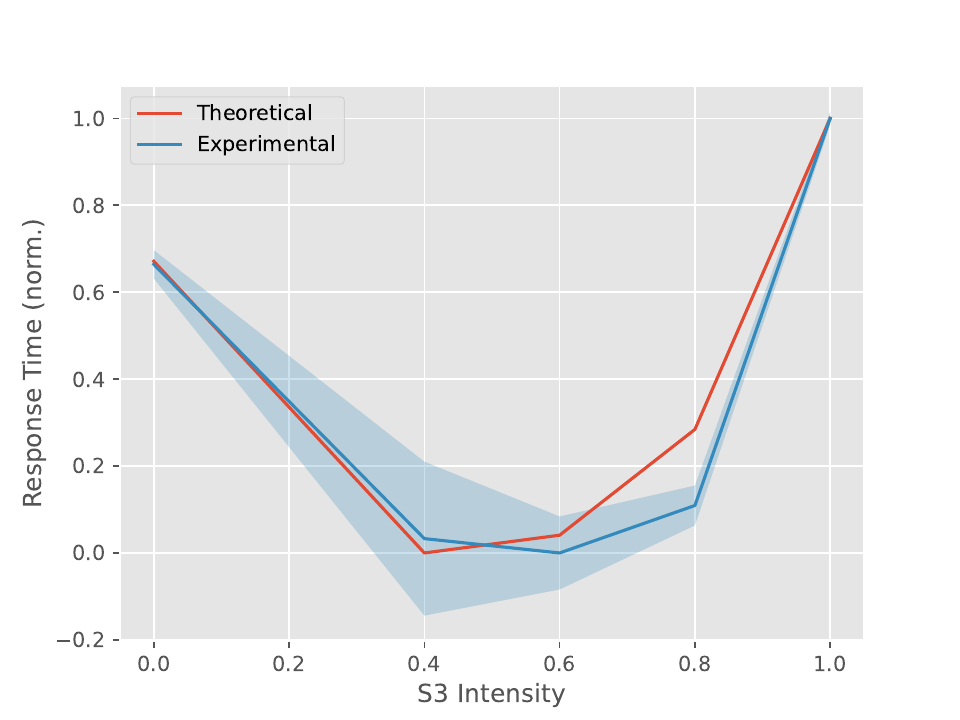}
            \caption{Normalized theoretical~\cite{pinciroli_performance_2023} and experimental request latency}
            \label{fig:ga-req-lat-norm}
        \end{subfigure}
        
        \\
        
        \begin{subfigure}{0.5\textwidth}
            \includegraphics[width=\textwidth]{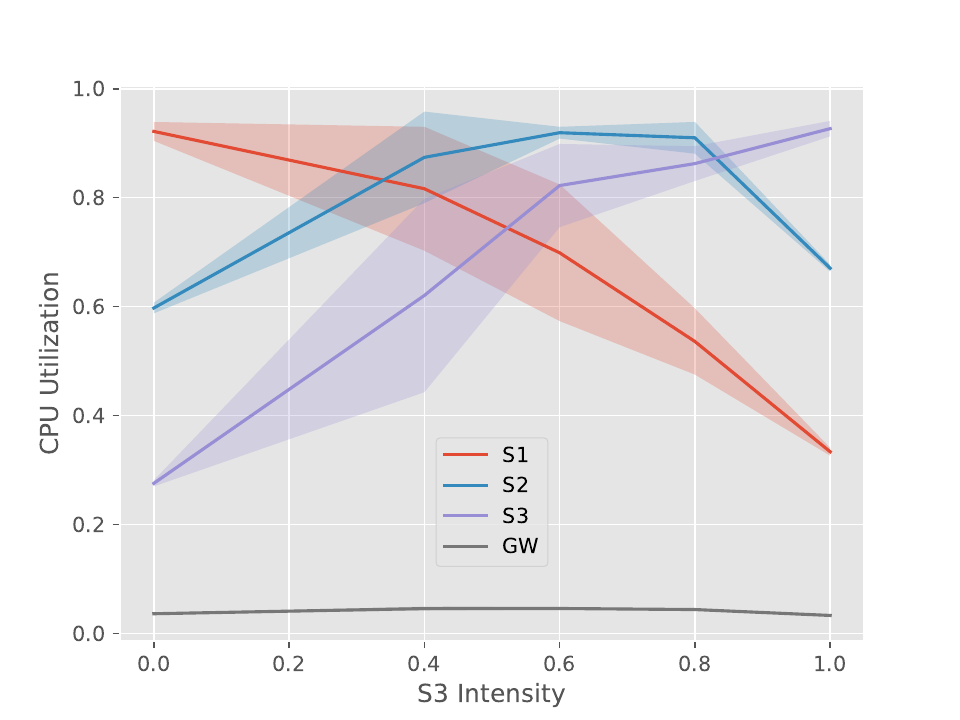}
            \caption{Experimental CPU utilization}
            \label{fig:ga-exp-cpu}
        \end{subfigure}

        & 
        
        \begin{subfigure}{0.5\textwidth}
            \includegraphics[width=\textwidth]{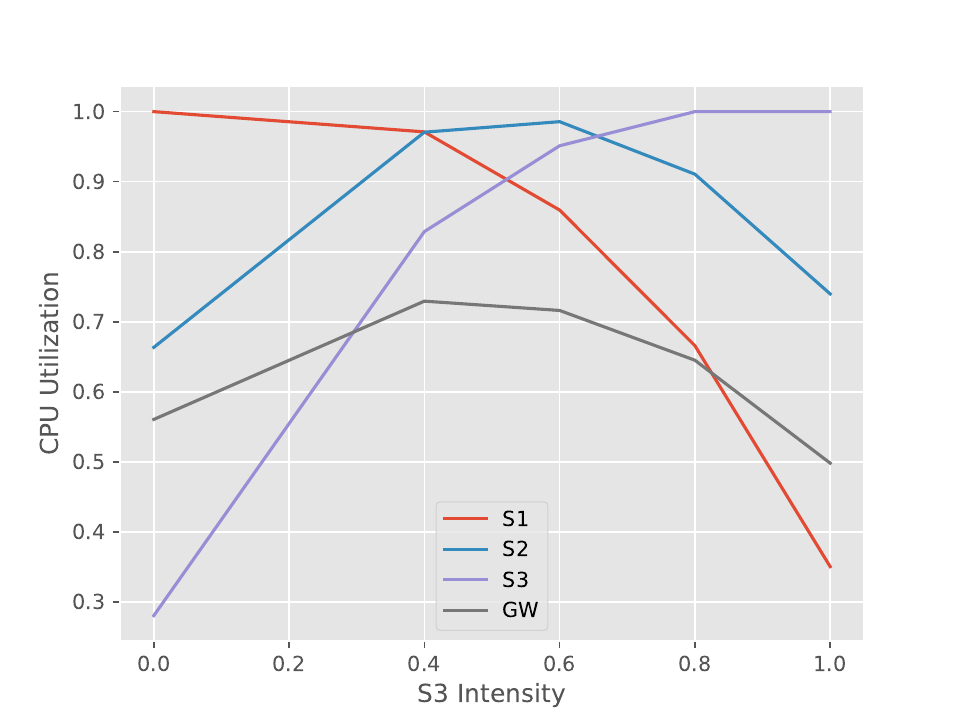}
            \caption{Theoretical CPU utilization~\cite{pinciroli_performance_2023}}
            \label{fig:ga-theo-cpu}
        \end{subfigure}

        \\

        \multicolumn{2}{c}{
            \begin{subfigure}{0.5\textwidth}
                \includegraphics[width=\textwidth]{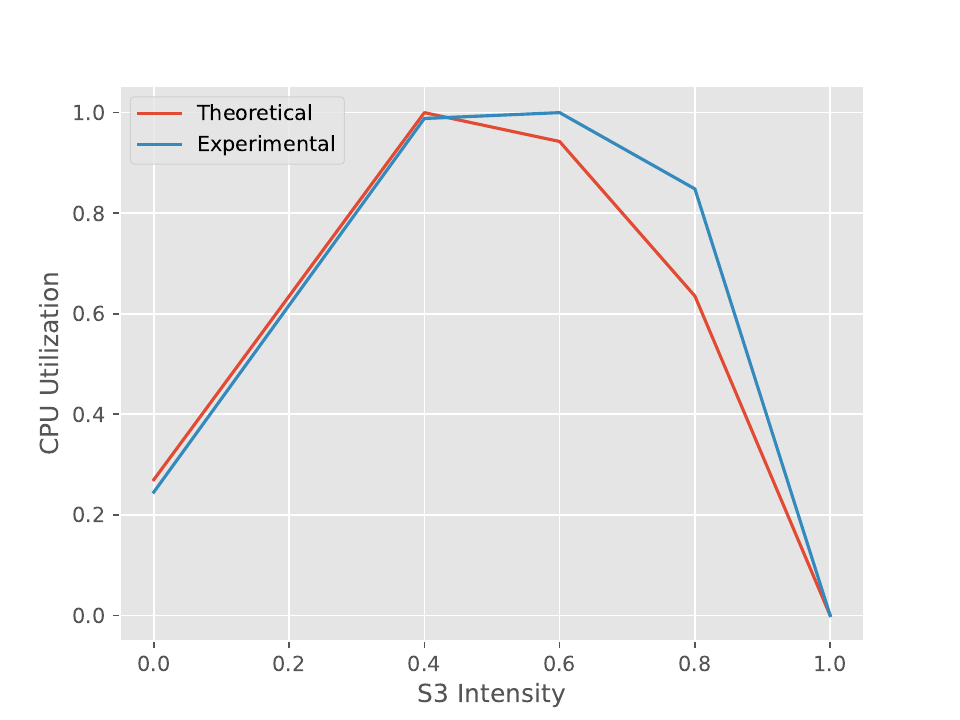}
                \caption{Comparison of the normalized theoretical~\cite{pinciroli_performance_2023} and experimental gateway service CPU utilization.}
                \label{fig:gateway-aggregator-gateway-results}
            \end{subfigure}
        }

    \end{tabular}
    \caption{Overview of the theoretical~\cite{pinciroli_performance_2023} and experimental results in terms of response time and CPU utilization of the \textit{gateway aggregation} pattern.}
    \label{fig:gateway-aggregator-results}
\end{figure}

\begin{table}
\scriptsize
    \centering
    \begin{tabular}{|c|c|c|c|}
        \hline
        \textbf{Metric} & \textbf{Spearman's $r$} & \textbf{MAE} & \textbf{Normalized MAE} \\
        \hline\hline
        \textit{Response Time} & $0.869^{*}$& $702$ms& $5.12\%$\\
        \hdashline
        \textit{CPU Utilization S1} & $0.931^{*}$ & $10.80\%$ & $10.66\%$\\
        \textit{CPU Utilization S2} & $0.782^{*}$ & $6.41\%$ & $8.87\%$\\
        \textit{CPU Utilization S3} & $0.934^{*}$ & $11.08\%$ & $9.42\%$\\
        \textit{CPU Utilization GW} & $0.871^{*}$ & $58.87\%$ & $12.20\%$\\
        \hline
    \end{tabular}
    \caption{Correlation and deviation metrics of the response time and CPU utilization evaluating the relationship between experimental observations and theoretical results of the \textit{gateway aggregation} pattern ($^{*} p < 0.001$).}
    \label{tab:results-gateway-aggregator}
\end{table}

    \rev{\emph{Comparative analysis with theoretical models.}}
    %
    The comparison between the experimental and the theoretical results of the gateway aggregator pattern shows that although their absolute values differ notably, they show similar behavioral patterns. 
    The graphs that are shown in Figures~\ref{fig:ga-req-lat-abs} and~\ref{fig:ga-req-lat-norm} exemplify this phenomenon as, on average, the absolute request delay is $\approx702$ milliseconds, while the normalized delay differs merely $5.12\%$ on average. 
    To reiterate, the increase in absolute difference is an expected side effect of the study design because a trade-off is made between system throughput and response delay \rev{(as described in Section~\ref{sec:load-amplification})}.
    Although the absolute differences are significant, a comparison between the theoretical and experimental system behaviors can still be made (i.e., their relative results, \rev{see Section~\ref{sec:analysis-metrics}}).
    The behavior expressed in the experimental service is very similar to the theoretical results, such that a heterogeneous load decreases message delay, suggesting that neither service 1 nor service 3 is a system bottleneck.
    Thus, the load is more or less equally distributed across the system.
    This similarity is supported by the statistical analysis (shown in Table~\ref{tab:results-gateway-aggregator}), which identified a significant and very strong correlation between the theoretical and experimental results ($r = 0.869, p < 0.001$).

    It stands out that none of the system services are CPU-bound (see Figures~\ref{fig:ga-exp-cpu} and~\ref{fig:ga-theo-cpu}) —- meaning that none of the services have $100\%$ CPU utilization.
    This is slightly inconsistent with the theoretical model, which estimates that the CPU utilization is higher.
    A notable observation is that the CPU utilization of the gateway aggregator is substantially lower than the theoretical model’s estimate, differing by $58.87\%$ on average. 
    This, however, can trivially be attributed to the study design because it uses a custom implementation of the gateway aggregator which does not execute a synthetic workload and thus performs substantially less work than the other services.
    Regardless, the results in Table~\ref{tab:results-gateway-aggregator} show that the theoretical and experimental results correlate strongly, as $r > 0.8$ for services 1, 3, and the gateway ($p < 0.001$), and $r = 0.78$ for service 2 ($p < 0.001$).
    For clarification, Figure~\ref{fig:gateway-aggregator-gateway-results} is added to visualize the similarity in behavior between the experimental and theoretical gateways.

    \rev{\emph{Empirical results.} Figure~\ref{fig:ga-req-lat-norm} assesses that the system latency is subject to variations due to the nature of incoming requests, the lowest values are obtained in the case of S3 intensive requests varying between 0.5 and 0.6. This implies that the ratio of heterogeneous requests is high, i.e., S1 intensive requests also circulate in the system and need to be handled. This performance trend is further explained in Figure~\ref{fig:ga-exp-cpu} where we do notice (in the very same case of S3 intensive requests varying between 0.5 and 0.6) that none of the services represent a system bottleneck, their utilization is lower than 100\%. Moreover, we do notice 
    the utilization of S3 increases when the S3 intensive requests augment in the system, and the opposite trend is observed for S1 whose utilization tends to decrease. Summarizing, our empirical results confirm that heterogeneous load is a key factor for a successful application of the \emph{gateway aggregation} pattern, given that the aggregated load is distributed among all the available services. 
    }

\subsubsection{Gateway offloading}
\label{sec:results-gateway-offloading}


\begin{figure}
    \centering
    \begin{tabular}{cc}
        \begin{subfigure}{0.5\textwidth}
            \includegraphics[width=\textwidth]{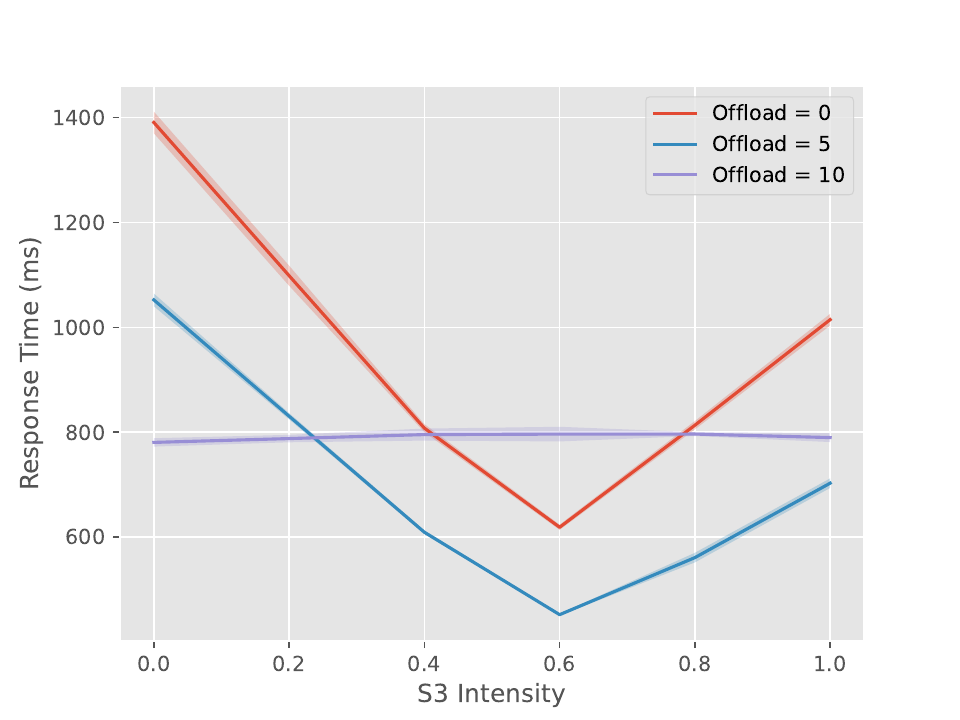}
            \caption{Experimental response delay}
            \label{fig:go-exp-req-lat}
        \end{subfigure}
        
        &
        
        \begin{subfigure}{0.5\textwidth}
            \includegraphics[width=\textwidth]{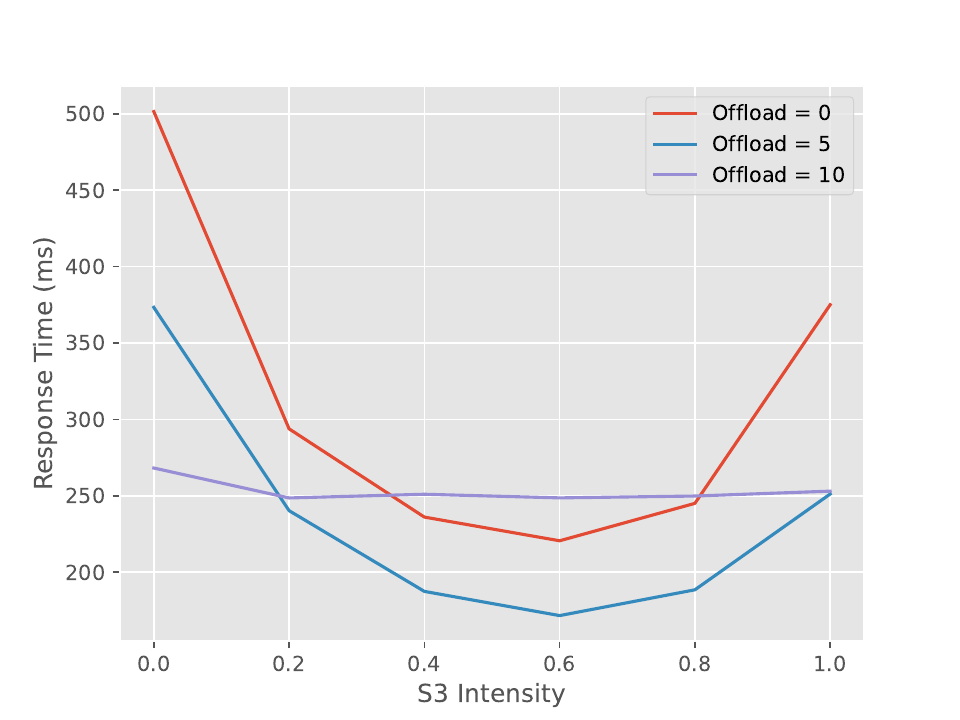}
            \caption{Theoretical response delay~\cite{pinciroli_performance_2023}}
            \label{fig:go-theo-req-lat}
        \end{subfigure}
    \end{tabular}

    \begin{tabular}{ccc}
        
        \begin{subfigure}{0.32\textwidth}
            \includegraphics[width=\textwidth]{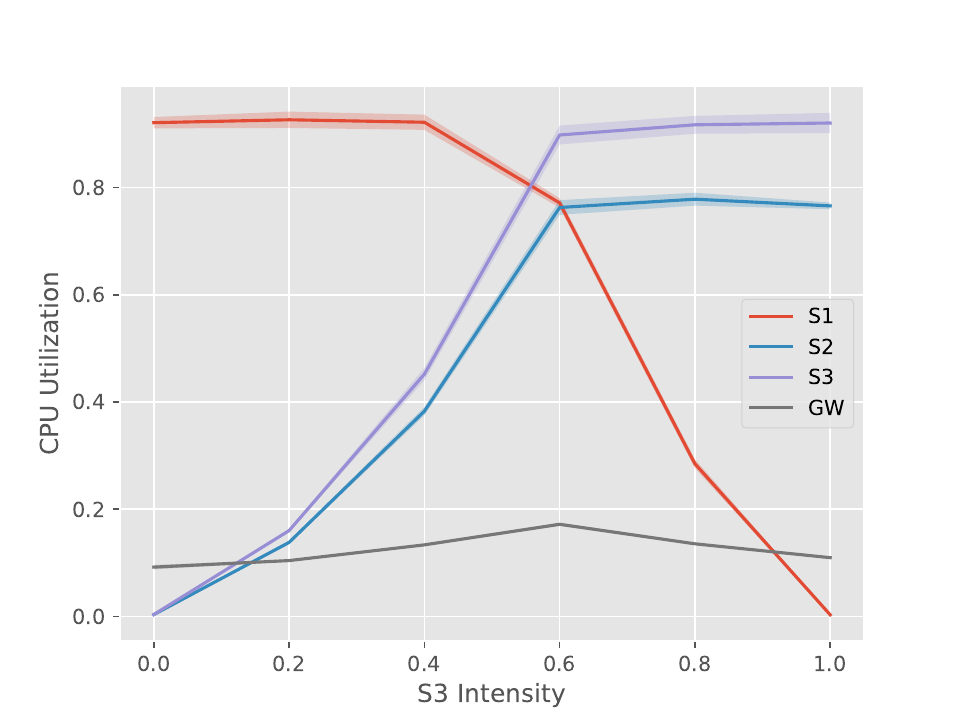}
            \caption{Experimental CPU utilization $\mbox{offload} = 0$}
            \label{fig:go-exp-cpu-offload-0}
        \end{subfigure}
        &
        \begin{subfigure}{0.32\textwidth}
            \includegraphics[width=\textwidth]{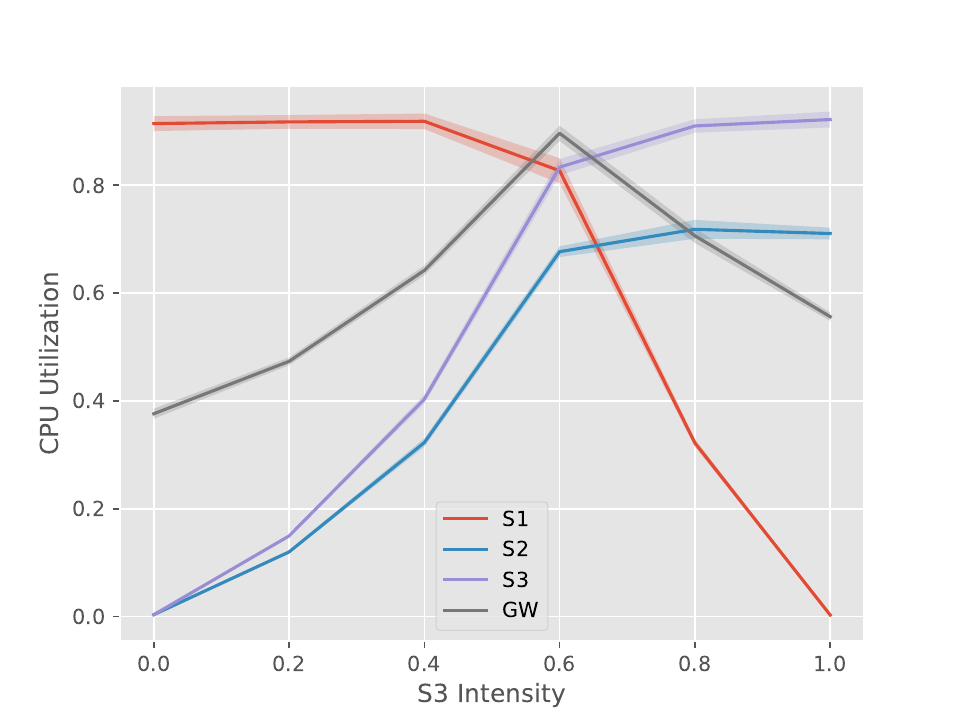}
            \caption{Experimental CPU utilization $\mbox{offload} = 5$}
            \label{fig:go-exp-cpu-offload-5}
        \end{subfigure}
        &
        \begin{subfigure}{0.32\textwidth}
            \includegraphics[width=\textwidth]{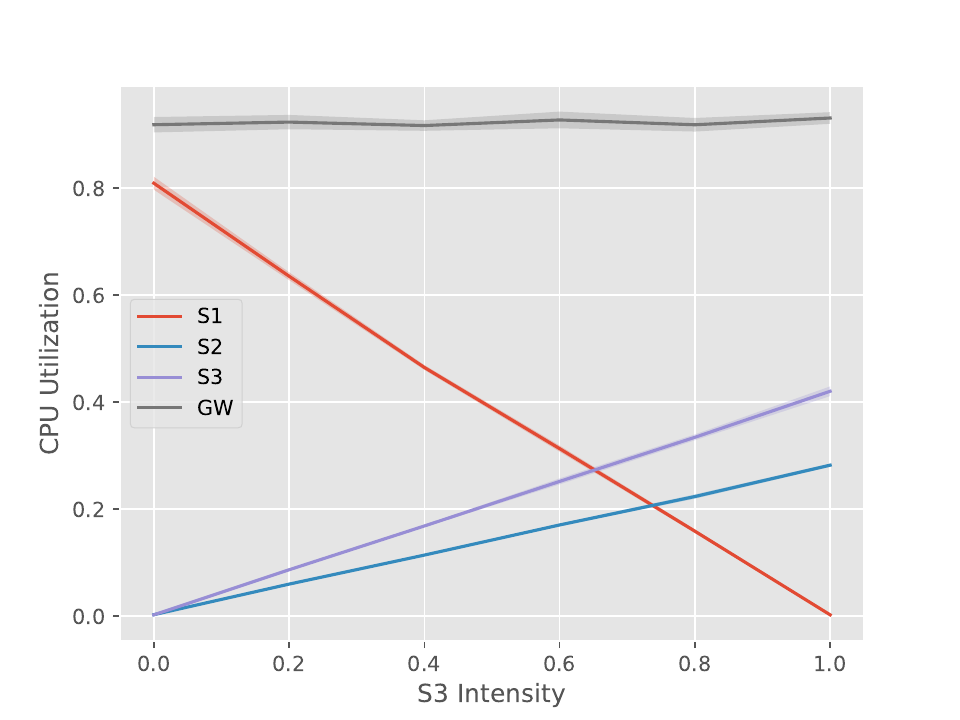}
            \caption{Experimental CPU utilization $\mbox{offload} = 10$}
            \label{fig:go-exp-cpu-offload-10}
        \end{subfigure}
        
        \\

        \begin{subfigure}{0.32\textwidth}
            \includegraphics[width=\textwidth]{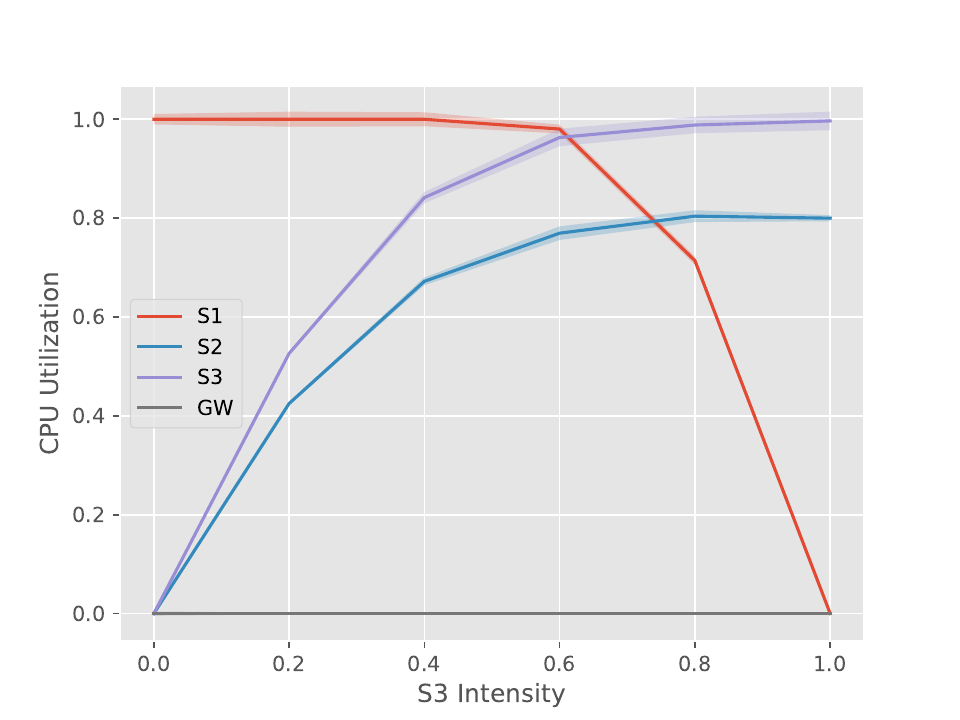}
            \caption{Theoretical CPU utilization $\mbox{offload} = 0$~\cite{pinciroli_performance_2023}}
            \label{fig:go-theo-cpu-offload-0}
        \end{subfigure}
        &
        \begin{subfigure}{0.32\textwidth}
            \includegraphics[width=\textwidth]{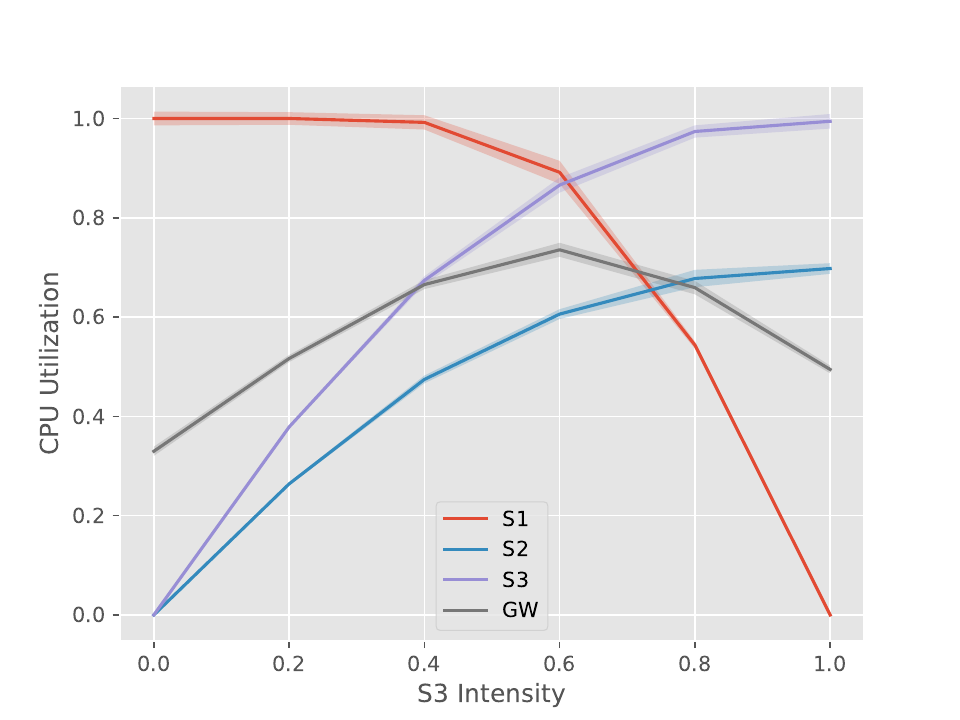}
            \caption{Theoretical CPU utilization $\mbox{offload} = 5$~\cite{pinciroli_performance_2023}}
            \label{fig:go-theo-cpu-offload-5}
        \end{subfigure}
        &
        \begin{subfigure}{0.32\textwidth}
            \includegraphics[width=\textwidth]{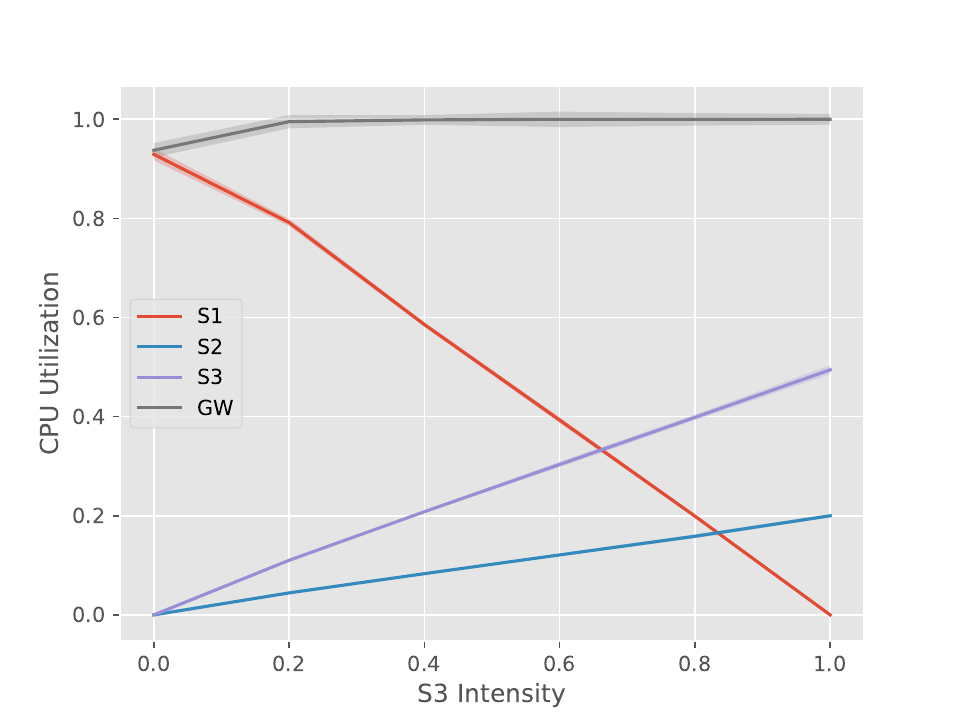}
            \caption{Theoretical CPU utilization $\mbox{offload} = 10$~\cite{pinciroli_performance_2023}}
            \label{fig:go-theo-cpu-offload-10}
        \end{subfigure}
        
    \end{tabular}
    \caption{Overview of the theoretical~\cite{pinciroli_performance_2023} and experimental results in terms of response time and CPU utilization of the \textit{gateway offloading} pattern using $\mbox{offload} \in \{0, 5, 10\}$.}
    \label{fig:gateway-offloading-results}
\end{figure}

\begin{table}[!t]
\scriptsize
    \centering
    \begin{tabular}{|c|c|c|c||c|c|c||c|c|c|}
        \cline{2-10}
        \multicolumn{1}{c|}{} & \multicolumn{3}{c||}{$\mbox{offload} = 0$} & \multicolumn{3}{c||}{$\mbox{offload} = 5$} & \multicolumn{3}{c|}{$\mbox{offload} = 10$} \\
        \hline
        \textbf{} & \textbf{} & \textbf{} & \textbf{Norm.} & \textbf{} & \textbf{} & \textbf{Norm.} & \textbf{} & \textbf{} & \textbf{Norm.} \\
        \textbf{Metric} & \textbf{$r$} & \textbf{MAE} & \textbf{MAE} & \textbf{$r$} & \textbf{MAE} & \textbf{MAE} & \textbf{$r$} & \textbf{MAE} & \textbf{MAE} \\
        \hline
        
        \textit{Resp.}  & $0.911^{*}$ & $645$ms & $12.93\%$ 
                        & $0.874^{*}$ & $466$ms & $10.23\%$ 
                        & $ins.$ & $538$ms & $79.79\%$ \\
        \hdashline
        \textit{Ut. S1} & $0.889^{*}$ & $14.53\%$ & $9.88\%$ 
                        & $0.866^{*}$ & $8.85\%$ & $4.16\%$ 
                        & $0.986^{*}$ & $8.66\%$ & $3.27\%$ \\
                        
        \textit{Ut. S2} & $0.932^{*}$ & $10.86\%$ &  $12.49\%$ 
                        & $0.952^{*}$ & $7.05\%$ & $9.50\%$ 
                        & $0.986^{*}$ & $4.12\%$ & $1.09\%$ \\
                        
        \textit{Ut. S3} & $0.932^{*}$ & $16.20\%$ & $12.65\%$ 
                        & $0.971^{*}$ & $11.20\%$ & $8.72\%$ 
                        & $0.986^{*}$ & $4.25\%$ & $1.58\%$ \\
        
        \textit{Ut. GW} & $ins.$ & $12.44\%$ & $41.52\%$ 
                        & $0.874^{*}$ & $6.38\%$ & $14.40\%$ 
                        & $ins.$ & $6.56\%$ & $84.52\%$ \\
        \hline
    \end{tabular}
    \caption{Correlation and deviation metrics of the \textit{response time} (\textit{Resp.}) and CPU utilization (\textit{Ut.}) that evaluate the relationship between experimental observations and theoretical results of the \textit{gateway offloading} pattern for different amounts of offloading ($^{*} p < 0.001$).}
    \label{tab:results-gateway-offloading}
\end{table}

    \rev{\emph{Comparative analysis with theoretical models.}}
    %
    Looking at the response delay results of the gateway offloading pattern (shown in Figures~\ref{fig:go-exp-req-lat} and~\ref{fig:go-theo-req-lat}), it stands clear that the theoretical results do not fully match the experimental ones.
    Zooming in on $\mbox{offload} = 0$ and $5$, we observe the same bottleneck switch from service 1 to services 2 and 3 (as visualized in Figures~\ref{fig:go-exp-req-lat} and~\ref{fig:go-theo-req-lat}). 
    However, the fashion in which this switch happens is much smoother in the theoretical model.
    The theoretical model follows a concave downward curve, whereas the experimental results change almost linearly.
    Regardless, on a grand scale, the same behavior is captured, such that for offloading 0 and 5, a strong correlation between the models is found ($r > 0.8, p < 0.001$; shown in Table~\ref{tab:results-gateway-offloading}).
    The models also capture the bottleneck behavior found when too much processing is offloaded, capturing the constant response rate perfectly.
    This conclusion cannot be drawn from any of the metrics shown in Table~\ref{tab:results-gateway-offloading}, as each of these becomes very sensitive to noise when the rate of change of a curve is zero, for which there is no apparent reason to refute this observation.
    The added benefit of offloading processing to a gateway is preserved in this experiment, as where $\mbox{offload} = 5$, the response time is uniformly lower than where $\mbox{offload} = 0$.

    Looking at the pattern’s CPU utilization (shown in Figures~\ref{fig:go-exp-cpu-offload-0} through~\ref{fig:go-theo-cpu-offload-10}), some interesting deviations from the theory can be observed.
    It stands out that none of the services, in none of the experiments, is CPU-bound, as was predicted by the theoretical model.
    Although similar bottlenecks and bottleneck switches can be observed, services never exceed $\approx90\%$ utilization; a gap that is most likely explained due to unaccounted-for inefficiencies of the system.
    At a larger scale, the behavior largely matches, which is complemented by statistical analysis, suggesting that all of the services’ theoretical and experimental results correlate strongly ($r > 0.8, p < 0.001$).
    Conversely, looking at Figures~\ref{fig:go-exp-cpu-offload-0} and~\ref{fig:go-theo-cpu-offload-0}, emphasizing $\mbox{offload} = 0$, it stands clear that the theoretical model fails to capture the CPU usage of the gateway.
    It predicted utilization of almost $0\%$ whereas the experimental results report it to lie between $10\%$ and $20\%$; processing power that is required for it to merely act as an intermediary without completing any real tasks.
    In addition, the theoretical model fails to capture the curvature of the gateway at this point too, as no correlation could be identified.
    Zooming in at $\mbox{offload} = 0$ and $5$, similar to the response time, the rate of change observed in the theoretical and experimental models differ quite notably in their concavity.
    This is somewhat prevalent for service 1, however, services 2, 3, and the gateway exemplify this phenomenon.
    Excluding when a service’s CPU utilization plateaus (e.g., when over $60\%$ of the requests are service 3 oriented), the curves’ concavity is mirrored, such that the theoretical model is upward concave and the experimental model is downward concave.
    Interestingly, where $\mbox{offload} = 5$, the CPU utilization of the gateway remains substantially higher compared to its theoretical counterpart, even exceeding that of services 1 and 3 and becoming the system's bottleneck; this reflects itself in the high normalized MAE score.

    \rev{\emph{Empirical results.} Figure~\ref{fig:go-exp-req-lat} illustrates the impact of the offloading procedure, and we can notice that the system response time decreases when the offloading of some computation is introduced (i.e., offload = 5). However, when offloading extensive computation (i.e., offload = 10) it turns out that the system response time does not vary, thus grasping that the offloading process can negatively impact the system performance. Figures~\ref{fig:go-exp-cpu-offload-0},~\ref{fig:go-exp-cpu-offload-5}, and~\ref{fig:go-exp-cpu-offload-10} support these findings while showing the CPU utilization of the gateway component and all other services. As expected, the CPU utilization of the gateway component is very low in the case of no offload (see Figure~\ref{fig:go-exp-cpu-offload-0}), since some runtime routines are executed only. The ratio of heterogeneous requests indeed contributes to a variation of the CPU utilization for all the system components in the case of introducing some offload (see Figure~\ref{fig:go-exp-cpu-offload-5}). The gateway component becomes the system bottleneck in case of pushing forward the offloading procedure (see Figure~\ref{fig:go-exp-cpu-offload-10}). In summary, our empirical results remark that offloading is not always beneficial for the system performance, it is strongly related to the type of operations that are offloaded, as well as the heterogeneity of incoming requests.}

\subsubsection{Pipes and filters}
\label{sec:results-pipes-and-filters}


\begin{figure}
    \centering
    \begin{tabular}{ccc}
    
        \begin{subfigure}{0.32\textwidth}
            \includegraphics[width=\textwidth]{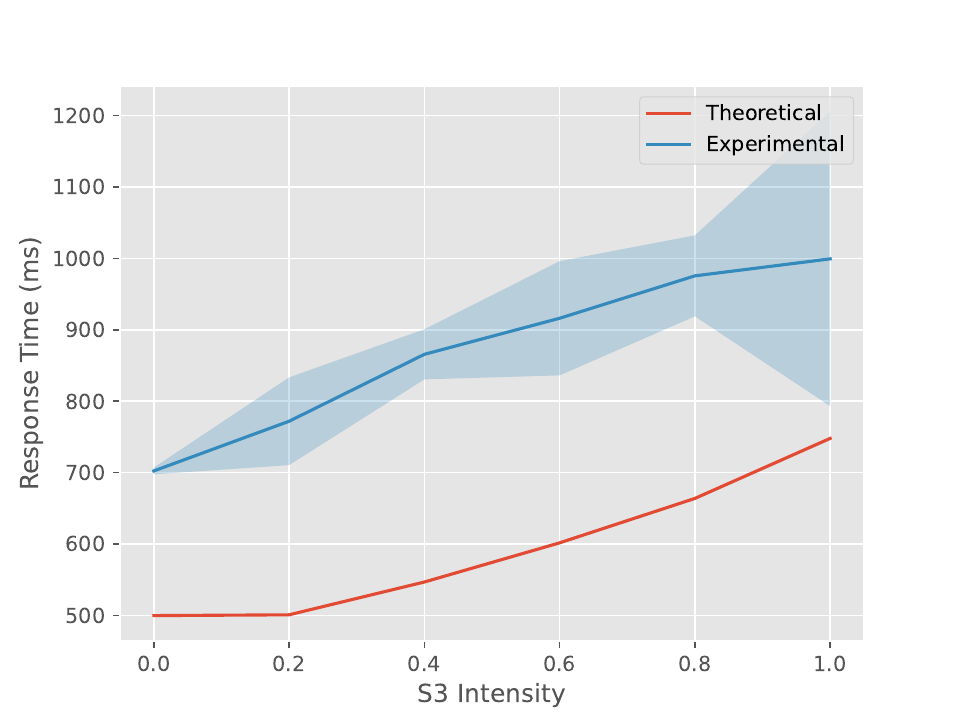}
            \caption{Response time joint pipeline with 1 CPU}
            \label{fig:pnf-1cpu-req_delay}
        \end{subfigure}
        &
        \begin{subfigure}{0.32\textwidth}
            \includegraphics[width=\textwidth]{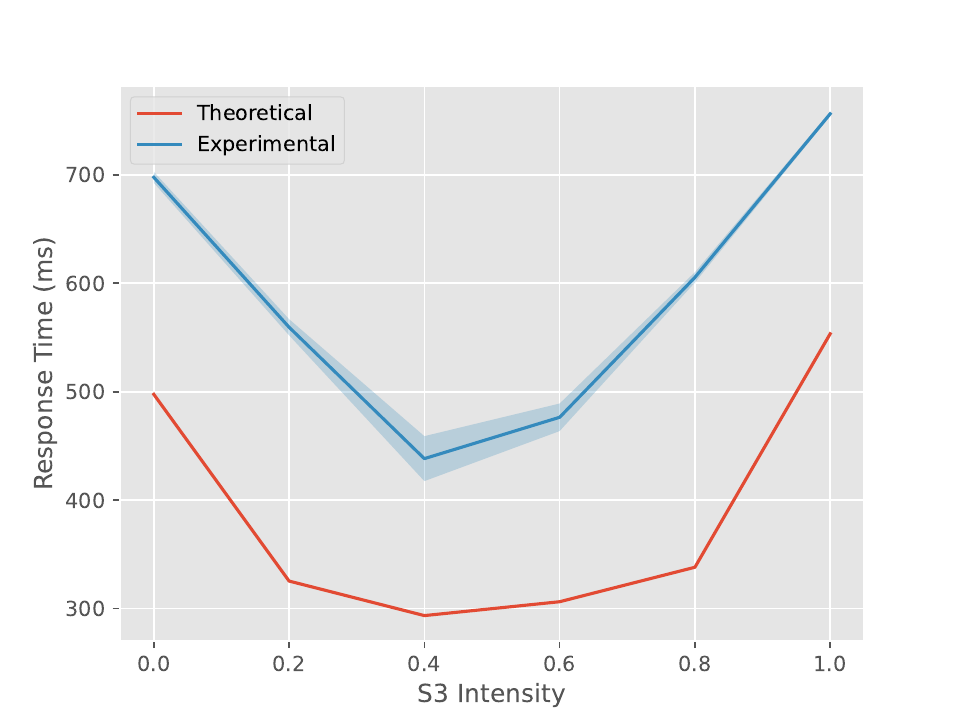}
            \caption{Response time joint pipeline with 2 CPUs}
            \label{fig:pnf-2cpu-req-delay}
        \end{subfigure}
        &
        \begin{subfigure}{0.32\textwidth}
            \includegraphics[width=\textwidth]{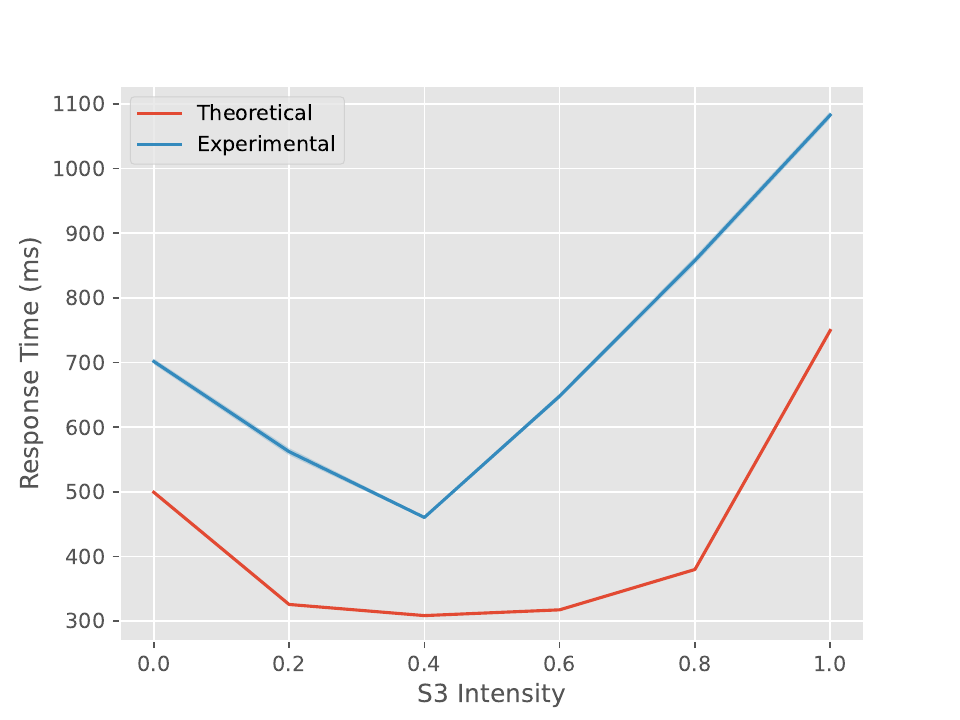}
            \caption{Response time separated pipelines}
            \label{fig:pnf-sep-req-delay}
        \end{subfigure}

        \\

        \begin{subfigure}{0.32\textwidth}
            \includegraphics[width=\textwidth]{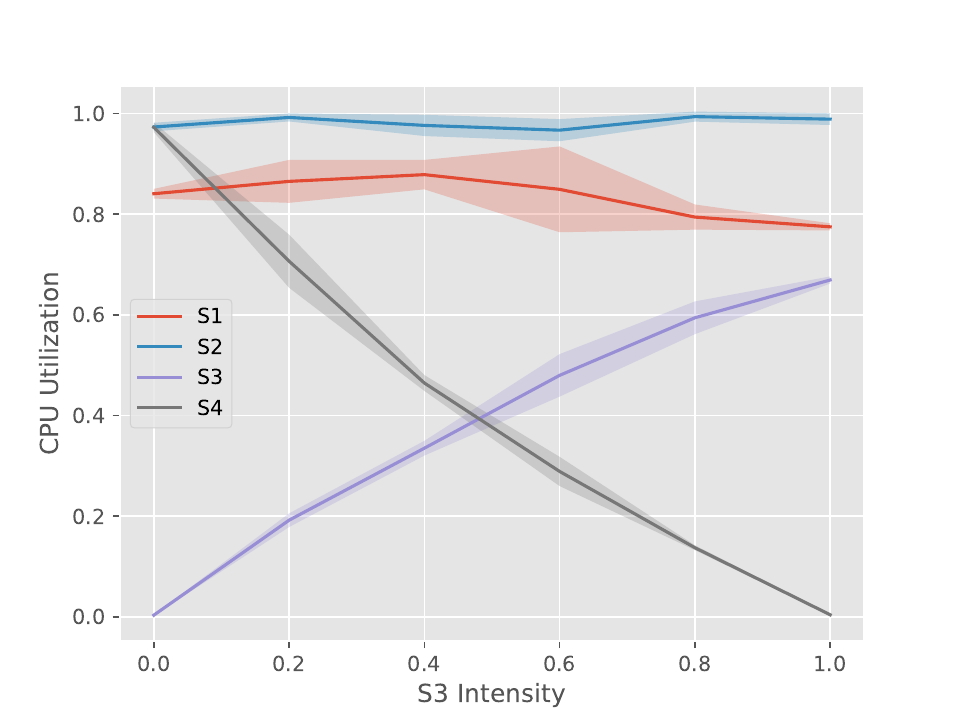}
            \caption{Experimental CPU utilization joint pipeline with 1 CPU}
            \label{fig:pnf-1cpu-exp-cpu}
        \end{subfigure}
        &
        \begin{subfigure}{0.32\textwidth}
            \includegraphics[width=\textwidth]{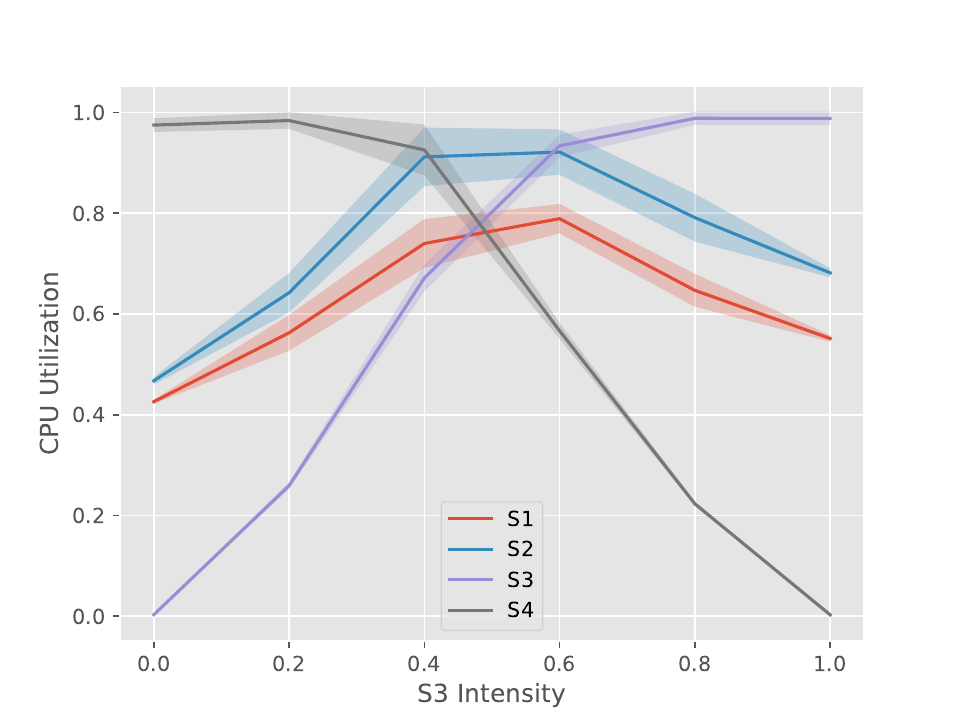}
            \caption{Experimental CPU utilization joint pipeline with 2 CPUs}
            \label{fig:pnf-2cpu-exp-cpu}
        \end{subfigure}
        &
        \begin{subfigure}{0.32\textwidth}
            \includegraphics[width=\textwidth]{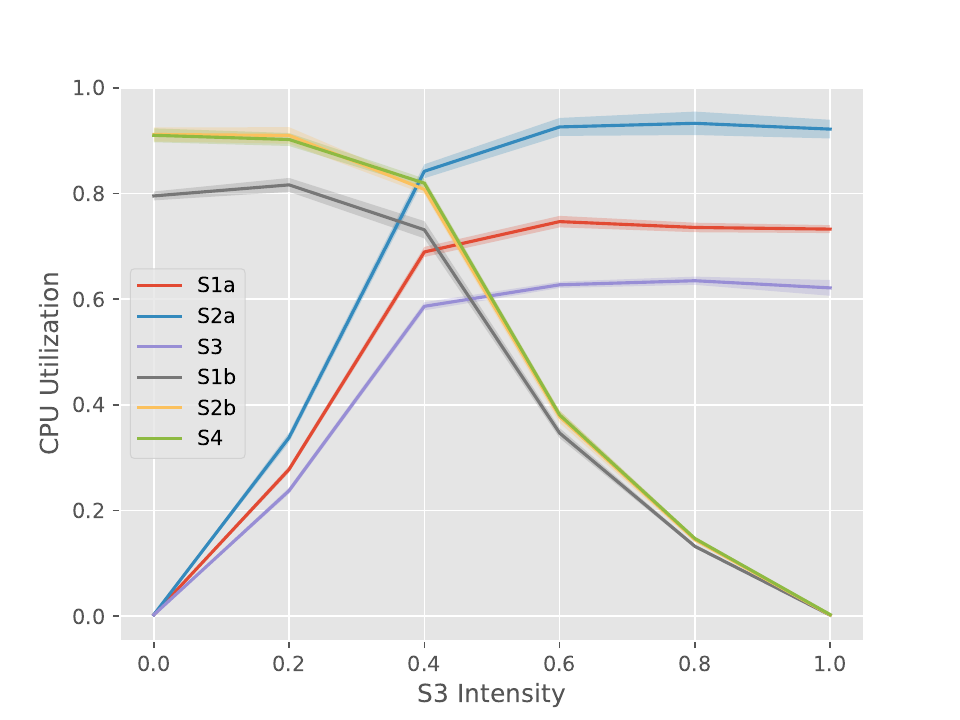}
            \caption{Experimental CPU utilization separated pipelines}
            \label{fig:pnf-sep-exp-cpu}
        \end{subfigure}
        
        \\

        \begin{subfigure}{0.32\textwidth}
            \includegraphics[width=\textwidth]{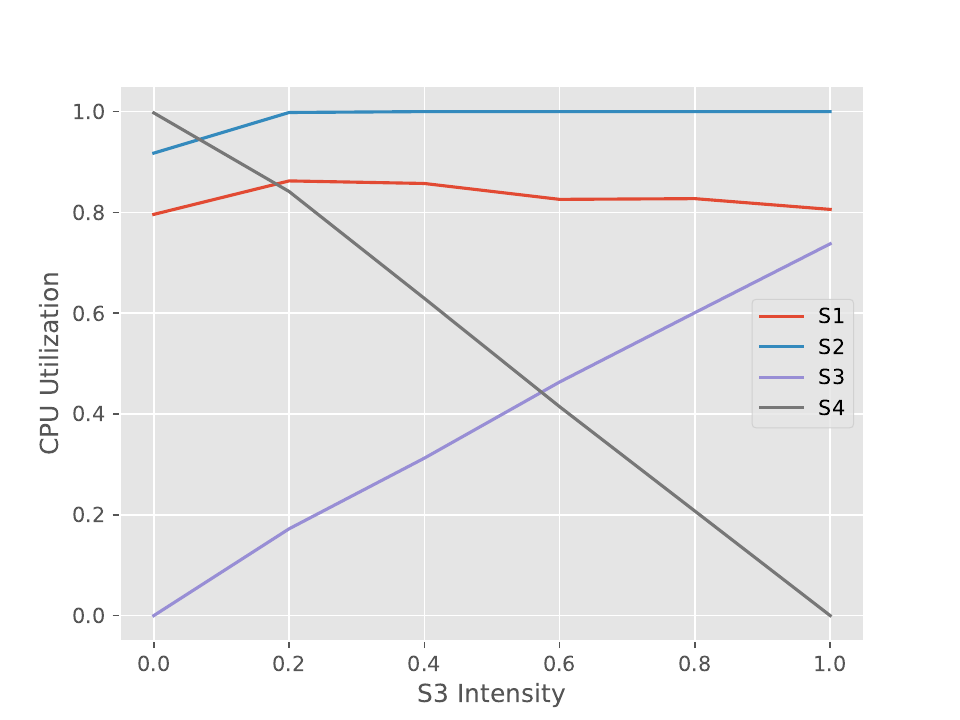}
            \caption{Theoretical CPU utilization joint pipeline with 1 CPU~\cite{pinciroli_performance_2023}}
            \label{fig:pnf-1cpu-theo-cpu}
        \end{subfigure}
        &
        \begin{subfigure}{0.32\textwidth}
            \includegraphics[width=\textwidth]{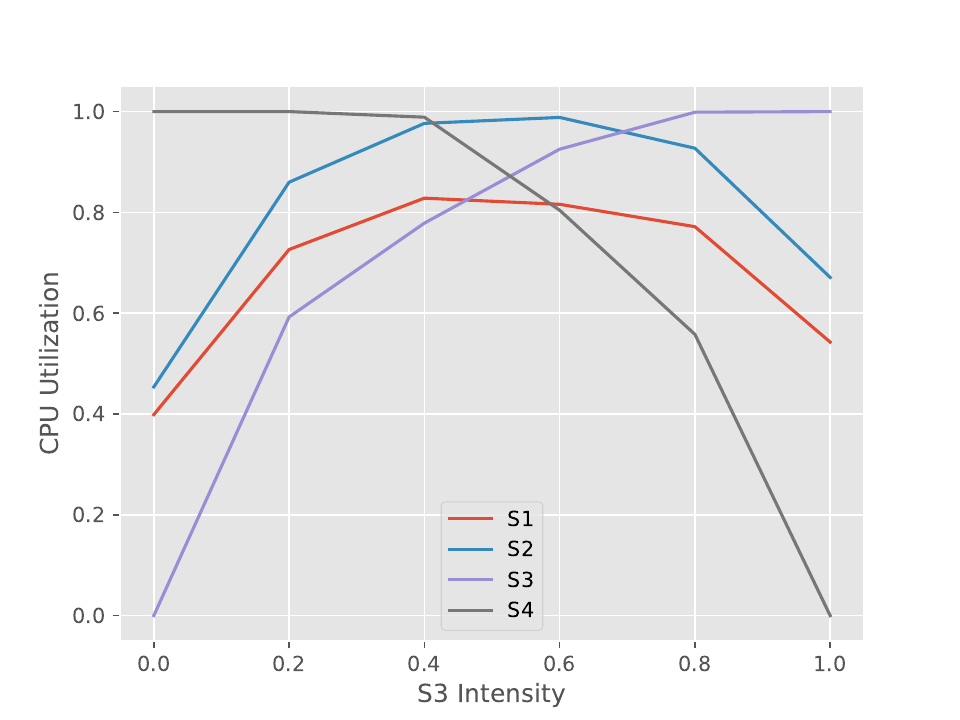}
            \caption{Theoretical CPU utilization joint pipeline with 2 CPUs~\cite{pinciroli_performance_2023}}
            \label{fig:pnf-2cpu-theo-cpu}
        \end{subfigure}
        &
        \begin{subfigure}{0.32\textwidth}
            \includegraphics[width=\textwidth]{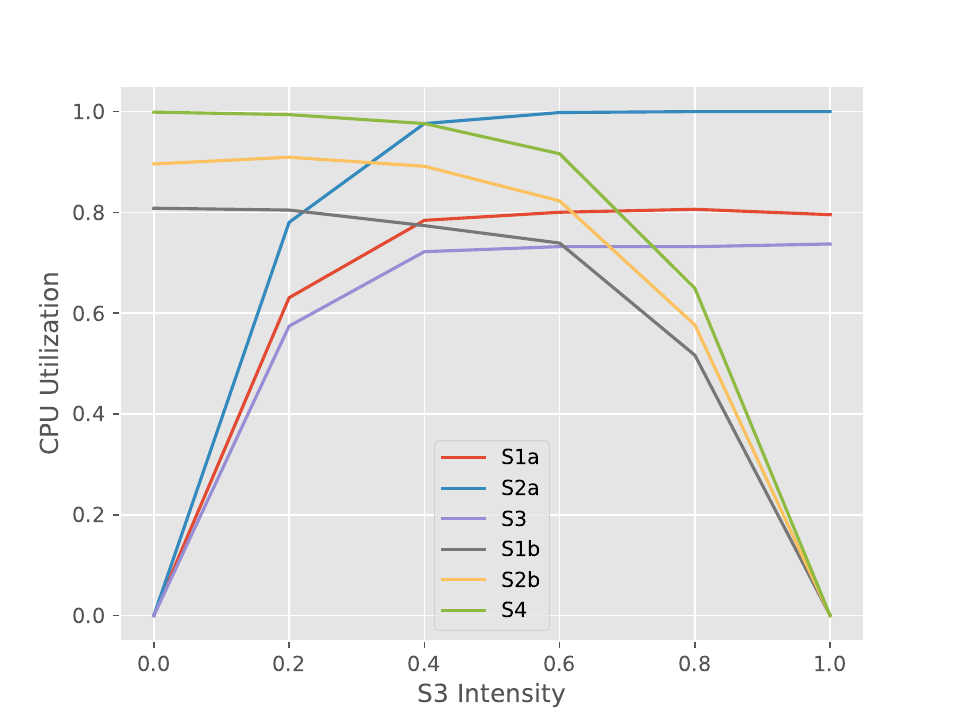}
            \caption{Theoretical CPU utilization separated pipelines~\cite{pinciroli_performance_2023}}
            \label{fig:pnf-sep-theo-cpu}
        \end{subfigure}
        
    \end{tabular}
    \caption{Overview of the theoretical~\cite{pinciroli_performance_2023} and experimental results in terms of response time and CPU utilization of the \textit{pipes and filters} pattern using 1 and 2 CPUs in the joint pipeline and using a separate pipeline.}
    \label{fig:pipes-and-filters-request-delay-results}
\end{figure}

\begin{table}[!ht]
\scriptsize
    \centering
    \begin{tabular}{|c|c|c|c||c|c|c||c|c|c|}
        \cline{2-10}
        \multicolumn{1}{c|}{} & \multicolumn{3}{c||}{\textit{Joint (1 CPU)}} & \multicolumn{3}{c||}{\textit{Joint (2 CPUs)}} & \multicolumn{3}{c|}{\textit{Separated}} \\
        \hline
        \textbf{} & \textbf{} & \textbf{} & \textbf{Norm.} & \textbf{} & \textbf{} & \textbf{Norm.} & \textbf{} & \textbf{} & \textbf{Norm.} \\
        \textbf{Metric} & \textbf{$r$} & \textbf{MAE} & \textbf{MAE} & \textbf{$r$} & \textbf{MAE} & \textbf{MAE} & \textbf{$r$} & \textbf{MAE} & \textbf{MAE} \\
        \hline
        
        \textit{Resp.}  & $0.780^{*}$ & $288$ms & $27.70\%$ 
                        & $0.982^{*}$ & $203$ms & $12.82\%$
                        & $0.874^{*}$ & $289$ms & $15.61\%$ \\
        \hdashline
        
        \textit{Ut. S1} & $0.391^{\diamond}$ & $3.72\%$ & $36.36\%$ 
                        & $0.907^{*}$ & $7.42\%$ & $14.92\%$ 
                        & $n/a$ & $n/a$ & $n/a$ \\
        
        \textit{Ut. S2} & $ins.$ & $2.25\%$ & $45.17\%$ 
                        & $0.938^{*}$ & $8.54\%$ & $13.09\%$ 
                        & $n/a$ & $n/a$ & $n/a$ \\
                        
        \textit{Ut. S3} & $0.986^{*}$ & $3.04\%$ & $4.91\%$ 
                        & $0.964^{*}$ & $8.12\%$ & $8.03\%$ 
                        & $0.817^{*}$ & $13.21\%$ & $8.47\%$ \\
        
        \textit{Ut. S4} & $0.986^{*}$ & $8.73\%$ & $7.91\%$ 
                        & $0.963^{*}$ & $11.34\%$ & $10.67\%$ 
                        & $0.969^{*}$ & $22.97\%$ & $18.21\%$ \\

        \hline\hline
        
        \textit{Ut. S1a}    & $n/a$ & $n/a$ & $n/a$ 
                            & $n/a$ & $n/a$ & $n/a$ 
                            & $0.888^{*}$ & $10.62\%$ & $8.36\%$ \\
        
        \textit{Ut. S2a}    & $n/a$ & $n/a$ & $n/a$ 
                            & $n/a$ & $n/a$ & $n/a$ 
                            & $0.881^{*}$ & $13.28\%$ & $9.03\%$ \\
        
        \textit{Ut. S1b}    & $n/a$ & $n/a$ & $n/a$ 
                            & $n/a$ & $n/a$ & $n/a$ 
                            & $0.933^{*}$ & $14.11\%$ & $17.86\%$ \\
        
        \textit{Ut. S2b}    & $n/a$ & $n/a$ & $n/a$ 
                            & $n/a$ & $n/a$ & $n/a$ 
                            & $0.957^{*}$ & $16.53\%$ & $18.25\%$ \\
        
        \hline
    \end{tabular}
    \caption{Correlation and deviation metrics of the response time (\textit{Resp.}) and CPU utilization (\textit{Ut.}) that evaluate the relationship between experimental observations and theoretical results of the \textit{pipes and filters} pattern for different amounts of offloading ($^{*} p < 0.001$; $^{\diamond} p < 0.05$).}
    \label{tab:results-pipes-and-filters}
\end{table}

    \rev{\emph{Comparative analysis with theoretical models.}}
    %
    Observing the 1-CPU pipes and filters experimental response delay (shown in Figure~\ref{fig:pnf-1cpu-req_delay}), the theoretical model gives an accurate representation of the bottleneck switch as both experiments show an almost linearly increasing pattern.
    This is corroborated by the Spearman correlation (shown in Table~\ref{tab:results-pipes-and-filters}), which suggests a strong correlation ($r = 0.78$ and $p < 0.001$).
    The theoretical model does seem to consistently underestimate the results slightly, which is visible from the curvature, which is concave upwards, whereas the experimental results suggest a concave downwards evolution.
    The CPU utilization corroborates this, showing a clear bottleneck in the shared tasks (see Figures~\ref{fig:pnf-1cpu-exp-cpu} and~\ref{fig:pnf-1cpu-theo-cpu}).
    It shows that service 2 is continuously CPU-bound, which is consistent with the theoretical model.
    However, when observing their correlation, we do not see a strong correlation, which is trivially explained because the CPU utilization does not change and thus the results are largely based on noise.
    However, these behaviors are very similar as the MAE score is only $2.25\%$. 
    Service 1 experiences the exact same phenomenon.
    Services 3 and 4 explain the linear increase of the message delay, showing a clear linear shift of the workload from one to the other.
    Then, because service 3 requires more time to handle its request, the message delay increases.
    Interestingly, although the CPU utilization of Service 3 only deviates $3.04\%$, that of Service 4 is consistently overestimated, to an average of $8.73\%$.
    This overestimation could explain the underestimate in response delay discussed earlier. 
    In general, it can be said that the theoretical model captures the experimental behavior well, albeit in an optimistic view of the system. 
    
    Looking at the 2-CPU experiment clearly shows different behavior, providing a clear picture of a bottleneck switch from service 4 to service 3. 
    The response delay (shown in Figure~\ref{fig:pnf-2cpu-req-delay}) curve then follows a clear U-shape.
    Although the theoretical results still strongly correlate with the experimental counterparts ($r = 0.982, p < 0.001$), the theoretical model, once again, underestimates the curvature of the behavior. The experimental results behave linearly, whereas the theoretical results behave concave downwards at the left and right halves of the curve.
    Consequently, the measured behavior deviates on average by $12.82\%$.
    This phenomenon is very similar to the observed behavior in the 1-CPU experiment.
    Here, again, the CPU utilization complements this behavior perfectly when observing the grand structure (shown in Figure~\ref{fig:pnf-2cpu-exp-cpu} and~\ref{fig:pnf-2cpu-theo-cpu}).
    A clear bottleneck switch can be observed from service 4 to service 3, both of which are CPU-bound when $60\%$ of the messages are of their respective types.
    In the intermediary $20\%$ (between $0.4$ and $0.6$) the shared components carry the largest burden, however, are never CPU-bound.
    This general similarity is corroborated by the performed statistical analysis, as the theoretical and experimental CPU usage of all services very strongly correlate (for all, $r > 0.9$ and $p < 0.001$).
    Similar to the 1-CPU experiment, the theoretical model overestimates the rate of change in CPU utilization, such that the theoretical results are concave downwards and the experimental results concave upwards.
    This holds for all of the services in this experiment (i.e., including the shared components), for which their MAEs lie between $8\%$ and $11\%$.
    
    Finally, when observing the separated \textit{pipes and filters} results, we see a response delay similar to the 2-CPU experiment.
    The change response delay clearly shows the bottleneck switch between services, following a V-shape (shown in Figure~\ref{fig:pnf-sep-req-delay}).
    The results correlated strongly  ($r = 0.874, p < 0.001$).
    However, here again, the theoretical model underestimates the behavior’s curvature, yielding an average deviation of $15.61\%$.
    Looking at the CPU utilization (shown in Figures~\ref{fig:pnf-sep-exp-cpu} and~\ref{fig:pnf-sep-theo-cpu}), the global picture is again very similar.
    However, contrary to the 2-CPU experiment, the U-/V-curves disappeared as separate pipelines were used, for which the observed change of CPU utilization is effectively monotonic; effectively, as the influence of noise is more apparent when components plateau or bottleneck.
    Here, again, the theoretical and experimental results strongly correlate (for all services, $r > 0.8$ and $p < 0.001$), and, again, the theoretical model overestimates the service’s CPU utilization following the same difference in curvature as the previous two experiments, resulting in a deviation between $10.62\%$ and $22.97\%$; a quite substantial difference.
    
    \rev{Combining these results shows that the theoretical model captures the global behavior of the patterns well as indicated by the strong positive correlation across models.}
    However, CPU utilization is over\-estimated  and message response times are under\-estimated in the non-extrem\-ities of the experiment; i.e., when a component becomes (or gets very close to being) CPU-bound.
    The theoretical model does this by differing strongly in its curvature.
    CPU utilization consistently differs in concavity.
    In response times, it differs in a complementing fashion, such that the theoretical curve is concave downwards whereas the experimental curve is almost linear.

    \rev{\emph{Empirical results.} Figure~\ref{fig:pnf-1cpu-req_delay} shows that, when considering joint pipeline services, the system performance may largely suffer. This behavior is further captured in Figure~\ref{fig:pnf-1cpu-exp-cpu} where it is evident that service S2 represents the system bottleneck, but also service S1 does manifest a high CPU utilization, hence the overall response time increases. The scenario changes when considering both the joint pipeline services that benefit from additional resources, and the separated pipelines. Specifically, Figure~\ref{fig:pnf-2cpu-req-delay} depicts that the heterogeneity of incoming requests plays a crucial role in minimizing the system response time. Again, this is mainly due to the CPU utilization of services for which we can visualize the bottleneck switch in Figure~\ref{fig:pnf-2cpu-exp-cpu}. With separated pipelines, we do notice a similar trend, i.e., the system response time decreases in the case of heterogeneous requests, the lowest value is observed when S3 intensity is equal to 0.4 (see Figure~\ref{fig:pnf-sep-req-delay}). The CPU utilization is subject to bottleneck switches, but half of the services do not overcome 80\%, thus generating fewer delays for some services (see Figure~\ref{fig:pnf-sep-exp-cpu}). Summarizing, we can conclude that the ratio of heterogeneous requests plays a key role for this pattern too. Moreover, the foreseen design solutions may largely affect the system performance, both the separation of services and the usage of additional hardware resources have been assessed to be beneficial.}

\subsection{Lessons learned}
\label{sec:lessons}

    Experimental and theoretical results of the \emph{gateway aggregation} pattern demonstrate notable differences in absolute values, yet they exhibit similar behavioral patterns. 
    Despite the absolute differences, the statistical analyses reveal a strong correlation between theoretical and experimental results, indicating the similarity in system behaviors at a high level, deviating between $5\%$ and $12\%$ on average.
    CPU utilization observations suggest that although the gateway aggregator's performance differs from the theoretical model due to the study's custom implementation and workload characteristics, the observed behavior remains similar.
    
    The experimental results of the \emph{gateway offloading} pattern showcase deviations from the theoretical model, particularly in response delay behaviors and CPU utilization.
    While the theoretical model predicts smooth transitions in bottleneck switches, experimental data reveals variations influenced by offloading parameters and system load.
    Observations suggest discrepancies between theoretical estimations and experimental outcomes, mainly for CPU utilization under maximal/minimal loads and bottleneck identification.

    Experimental findings indicate clear bottleneck transitions and behavioral patterns in the \emph{pipes and filters} architecture, mainly for CPU utilization and response delay.
    The theoretical model generally captures the global behavior of patterns but tends to overestimate CPU utilization and underestimate message response times, especially when components approach CPU-bound states, deviating as much as $18\%$ on average.
    
    Notable deviations are observed between theoretical predictions and experimental results, highlighting the importance of refining theoretical models to better align with real-world scenarios.
    Overall, the results highlight the complexity of evaluating architectural patterns and emphasize the need for complementing theoretical models with empirical data.
    %
    \rev{As a guideline for software engineers, we recommend beginning by familiarizing themselves with the theoretical models that describe architectural patterns to understand the underlying principles and assumptions. 
    This can help save time and computing resources for predicting the performance behavior of microservice architectures.
    However, it is 
    necessary to validate theoretical models with empirical data through experimentation.
    Real-world implementations often deviate from theoretical models due to various factors such as custom implementation, workload characteristics, and system constraints.
    Thus, experimental evaluation on real hardware is needed to identify finer-grained system performance behaviors.
    These experimental findings can in turn be used to improve the theoretical models, 
    e.g., adjusting assumptions, incorporating additional variables, or refining the analyzed algorithms, to better align with observed behaviors.}

    \rev{
    
    \subsubsection{Lessons for software architects}
    \label{sec:lessons-for-architects}
        
        Design patterns come with inherent downsides and risks, many of which are outlined in Microsoft's cloud design pattern index~\cite{microsoft2022cloud}.
        In terms of performance, all three patterns discussed in this work introduce the risk of a performance bottleneck.
        In the \textit{gateway aggregation} and \textit{gateway offloading} patterns, there's a risk that the gateway component could become a bottleneck due to the introduction of an additional service in the system.
        The \textit{pipes and filters} pattern can also introduce a bottleneck, since 
        the performance of a pipeline is dependent on its slowest operation, limiting the performance of other steps.
        Our study validates the findings presented by Pinciroli et al.~\cite{pinciroli_performance_2023}, identifying bottleneck switches in the three design patterns while considering the impact of heterogeneous workload.
        

        When observing the \emph{gateway offloading} pattern, we see that the CPU utilization of the gateway service piques for highly heterogeneous loads.
        In the case of homogeneous loads, the bottleneck switches to the backend services proxies by the gateway.
        This effect is amplified by the amount of work offloaded to the gateway.
        Although the gateway becomes a bottleneck, its high CPU consumption also indicates a large utilization of the services it proxies.
        This is reflected in the system's response times, which are significantly lower for highly heterogeneous loads, reaching an optimum when both the gateway and the backend services are highly utilized.  
        
        The bottleneck switch in the \emph{gateway aggregator} pattern is very similar.
        However, the gateway does not become the system's bottleneck at any point, even if 
        its utilization increases for heterogeneous loads. 
        This suggests that as long as the overhead of unpacking requests and repacking responses does not exceed the processing time of backend services, the added risk of this pattern is low.
        Intuitively, the system's bottlenecks are the emphasized services under homogeneous loads (e.g., service 1 is the bottleneck when most of requests are service 1 intensive).
        However, the results highlight non-trivial performance behavior under heterogeneous loads.
        Here, the service that bottlenecks is never the slowest task in a single user request.
        Because it has a medium computation time for every type of request, it becomes overwhelmed with high heterogeneity, while the other services run efficiently.
        This suggests that latent relationships among services should be accounted for during the design process of this pattern.
        
        Therefore, software architects, in charge of designing systems implementing the gateway aggregation or gateway offloading patterns, should closely account for the relationships between services.
        For example, by allocating resources to the gateway that are i) proportional to the resources allocated to the backend services, and ii) the number of requests that the backend services receive, i.e., the workload heterogeneity.
        
        Our results also highlight a bottleneck switch in the \textit{pipes and filters} pattern.
        In the joint pipeline (i.e., two separate pipelines share services) the observed effect is very similar to the gateway offloading pattern.
        Our analysis shows the impact of allocating resources not proportional to those of the following backend services (all services were allocated 1 CPU).
        When insufficient resources are allocated, the shared microservices are continuously CPU-bound, regardless of the heterogeneity of the workload.
        This bottleneck was resolved by allocating additional resources to the shared services.
        However, this allocation also introduces a bottleneck switch similar to the gateway patterns'.
        For highly heterogeneous loads the shared services become the system's bottleneck, accompanied by a reduced response time due to high utilization of the overall system.
        The separated pipeline (i.e., without shared tasks) intuitively does not suffer from the same bottleneck switch as the workload targets a different service.
        Instead, the shared pipeline highlights the ``slowest task in the pipeline'' more clearly.

        Hence, software architects should closely account for relationships between services, ensuring resource allocation is proportional to backend services and workload heterogeneity.
        For the pipes and filters pattern, consider the rate of change in utilization across services in a single pipeline, ensuring maximum utilization is not limited by the slowest task.
        Mitigate underutilization of pipelines for homogeneous loads to prevent resource wastage. 
        Further, pipelines become visibly underutilized for homogeneous loads, ultimately wasting resources.

    \subsubsection{Lessons for performance engineers}
    
        A large part of this study addresses microservice benchmark modeling.
        Our work employed $\mu$Bench~\cite{detti_bench_2023} that is a synthetic benchmarking tool for microservice architectures, offering great flexibility in service topology and resource requirements.
        We adjusted this system to increase its flexibility, specifically regarding message-dependent microservice behaviors and deploying real services alongside synthetic ones.
        Although our study targets a relatively small-scale implementation of architectural design patterns, we expect such a system to function well on larger-scale solutions too.
        Studies on these systems can address other important performance metrics like network delays or system scalability, especially when support for non-synthetic services is increased (e.g., event-driven architectures using Kafka or concrete databases like Cassandra).
        This would be a prerequisite when evaluating some of the microservice design patterns not addressed in this study and when evaluating them at scale.

        The results of this study highlighted the similarities and differences between experimental and theoretical estimates of runtime performance behaviors (CPU utilization and response delay) of microservice design patterns under heterogeneous loads.
        By comparing the queuing network models introduced by Pinciroli et al.~\cite{pinciroli_performance_2023}, with benchmarks on concrete implementations of these architectures, we observe many similarities between performance estimates on a global scale.
        However, our results also indicate a systematic discrepancy. 
        The theoretical models overestimate resource utilization and consequently underestimate request delays, especially when the heterogeneity of the requests changes.
        Committing to these models should therefore be done with some care, especially in systems where request heterogeneity is expected to rapidly change.
        Not doing so could result in unnecessary under- and over-utilization of computing resources.

        To summarize, performance engineers should be aware that QN models are a useful abstraction for predicting the performance of microservice-based architectures.
        However, the experimental evaluation on real hardware is needed to identify finer-grained behaviors.
        The experimental results can in turn be used to improve the theoretical models, creating a continuous feedback loop. 
        
        
    }

\subsection{Threats to validity}
\label{sec:threats}

    Besides inheriting all limitations of design patterns and integrating software performance engineering methods and practices~\cite{5975176, bondi2016incorporating}, 
    our work exhibits the following threats to validity~\cite{Wohlin2012}.
    
    \paragraph{External threats} 
    The study acknowledges limitations regarding the generalization of results due to its focus on only three design patterns and specific conditions such as heterogeneous workloads and operational profiles leading to bottleneck switches.
    While efforts were made to address these limitations by considering large variations in workloads, the study recognizes the need to demonstrate the impact of design patterns across different real-world applications.
    To address this, we plan to conduct further experiments with design patterns in various industrial applications, potentially spanning multiple domains.

    \paragraph{Internal threats} 
    As part of the study design, a trade-off was made between the number of requests and the amount of computing necessary to handle a single request.
    Systems that do not make this trade-off might yield different results.
    Beyond an absolute difference in response delay (which some of our experiments showed to be relatively high), we think this trade-off to have minimal impact on the study results. Because it is analogous to a client that merges multiple API requests into one, it does not affect bottleneck behavior.
    In addition, some manual experimentation was performed to fine-tune this trade-off, ensuring that the system is not systematically over- or underutilized.
    
    A second internal threat to the performed experimentation is the locality of load generators and the microservices as they were run in the same compute cluster.
    \rev{It is worth remarking that this design choice may introduce a discrepancy between experimental conditions and real scenarios.
    In real-world deployments, systems may be distributed across different physical locations or networks, leading to varied network conditions and potential latency issues that are not accurately represented in our experiments. 
    We recognize that this design decision results in an approximation of the response time compared to a real system, which might experience additional delays. This could moderately impact the system's efficiency, leading to somewhat optimistic results. However, it is improbable that this would lead to misrepresent the behavior of design patterns.
    }

    Finally, the experiments used a custom implementation of the gateway aggregation in the respective experiments because $\mu$Bench~\cite{detti_bench_2023} did not support this behavior natively.
    A consequence is that the workload of this system is not amplified like the other microservices, for which the reported CPU utilization is relatively low.
    This does not impact the reported results as the gateway was not the system's bottleneck.
    In addition, the results showed that regardless of the workload, the gateway's expected behavior was preserved.

    \paragraph{Construct threats} 
    Software benchmark results are commonly affected by some amount of noise due to background processes or noisy neighbors.
    To account for this, each experiment was repeated 6 times, thus keeping under control the accuracy of presented measurements.
    
    Comparing the experimental and theoretical results is by no means trivial as both systems are strongly affected by their configurations, and these are not trivially translated across systems.
    It is important to note that the primary goal of this study is to capture the \textit{behavior} of the design patterns rather than absolute differences.
    To increase resilience against configuration differences, this study employs two scale-invariant metrics in tandem with one absolute metric to compare the experimental and theoretical results.
    In addition, some amount of experimentation was performed to ensure that the experimental environment was very similar to its theoretical counterpart.
    Here, extra attention was spent on the artificial workload amplification, as anticipated in \textit{internal threats} (also argued in Section~\ref{sec:microservice-topology-modeling}).
    It is worth remarking that there exists a fairly large window in which the experimental and theoretical results behave very similarly, hence used values did not affect the results to any realistic extent.

\section{Conclusion and future work}
\label{sec:conclusion}

In this paper, we collect performance measurements to evaluate the characteristics of three design patterns that can be used in microservice systems under heterogeneous loads.
The main findings are: i) the experimental results confirm the trend observed by previous theoretical performance estimations, but ii) the absolute values of real measurements w.r.t. model-based predictions can deviate when the system is under minimal/maximal load, or the requests' heterogeneity implies a different utilization of resources.
Performance-based experimental results complement theoretical data by supporting software architects in understanding the impact of heterogeneous workloads on the performance behavior of design patterns.
We conclude that it is important to create a continuous feedback loop between theoretical models and empirical findings, which allows for iterative refinement.
This approach can significantly improve the accuracy and reliability of performance models over time.

Our future work research agenda includes addressing all the limitations discussed as part of threats to validity.
Moreover, we plan to further investigate the impact of design patterns in diverse systems\rev{, even considering different metrics of interest. For instance, the impact of design patterns on the scalability or the energy consumption of these systems is a promising research direction we intend to pursue in the near future}.
Empirical studies involving different stakeholders (mainly software architects and industrial practitioners) will be conducted to understand how they perceive our findings as support to quantify the impact of design patterns on the performance evaluation of microservice-based systems.


\section{Acknowledgements}

The authors would like to thank the anonymous reviewers for their precious feedback.  
This work has been partially funded by the MUR-PRIN project DREAM (20228FT78M), the MUR-PRO3 project on Software Quality, the MUR-PNRR project VITALITY (ECS00000041), the Wallenberg AI, Autonomous Systems and Software Program (WASP) funded by the Knut and Alice Wallenberg Foundation, and Australian Research Council grant no. DP210100041.
%
%

\bibliographystyle{elsarticle-num}
\bibliography{new_bibliography}





\end{document}